\documentclass[12pt]{article}
\usepackage{epsfig,epic,eepic}
\usepackage{amsfonts,amsmath}
\usepackage{hyperref}
\usepackage{hangcaption}

\numberwithin{equation}{section}

\newcommand{\smallfrac}[2] {{\textstyle{\frac{#1}{#2}}}}
\newcommand{\tinybox}[1] {\mbox{\tiny {#1}}}
\newcommand{\s}[1] {{\scriptstyle{#1}}}

\newcommand{\eqneq}{\hspace{-6pt}=\hspace{-6pt}}
\newcommand{\tr} {\mbox{tr}}
\newcommand{\diag} {\mbox{diag}}
\newcommand{\offdiag} {\mbox{offdiag}}
\newcommand{\N} {{\cal N}} 
\newcommand{\Z} {{\mathbb Z}} 
\newcommand{\One} {{\bf 1}} 

\newcommand{\Ident} {{\cal I}}
\newcommand{\Trace} {{\cal T}}
\newcommand{\phicl}[1][] {{\phi^{\text{c}}_{#1}}}

\newcommand{\epsFig}[2] {\raisebox{#1}{\psfig{figure=#2.eps}}}

\newcommand{\PFund} {\hspace{-6pt}\raisebox{0pt}{\setlength{\unitlength}{0.00016667in}
\begingroup\makeatletter\ifx\SetFigFont\undefined%
\gdef\SetFigFont#1#2#3#4#5{%
  \reset@font\fontsize{#1}{#2pt}%
  \fontfamily{#3}\fontseries{#4}\fontshape{#5}%
  \selectfont}%
\fi\endgroup%
{\renewcommand{\dashlinestretch}{30}
\begin{picture}(4908,123)(0,-10)
\thicklines
\path(54,54)(2754,54)
\blacken\path(2154.000,-96.000)(2754.000,54.000)(2154.000,204.000)(2154.000,-96.000)
\path(2154,54)(4854,54)
\end{picture}
}
}\hspace{-6pt}}
\newcommand{\PFundI} {\hspace{-6pt}\raisebox{0pt}{\setlength{\unitlength}{0.00016667in}
\begingroup\makeatletter\ifx\SetFigFont\undefined%
\gdef\SetFigFont#1#2#3#4#5{%
  \reset@font\fontsize{#1}{#2pt}%
  \fontfamily{#3}\fontseries{#4}\fontshape{#5}%
  \selectfont}%
\fi\endgroup%
{\renewcommand{\dashlinestretch}{30}
\begin{picture}(2508,123)(0,-10)
\thicklines
\path(54,54)(2454,54)
\end{picture}
}
}\hspace{-6pt}}
\newcommand{\PFundA} {\hspace{-6pt}\raisebox{0pt}{\setlength{\unitlength}{0.00016667in}
\begingroup\makeatletter\ifx\SetFigFont\undefined%
\gdef\SetFigFont#1#2#3#4#5{%
  \reset@font\fontsize{#1}{#2pt}%
  \fontfamily{#3}\fontseries{#4}\fontshape{#5}%
  \selectfont}%
\fi\endgroup%
{\renewcommand{\dashlinestretch}{30}
\begin{picture}(2508,123)(0,-10)
\thicklines
\dashline{360.000}(54,54)(2454,54)
\end{picture}
}
}\hspace{-6pt}}
\newcommand{\PFundB} {\hspace{-6pt}\raisebox{0pt}{\setlength{\unitlength}{0.00016667in}
\begingroup\makeatletter\ifx\SetFigFont\undefined%
\gdef\SetFigFont#1#2#3#4#5{%
  \reset@font\fontsize{#1}{#2pt}%
  \fontfamily{#3}\fontseries{#4}\fontshape{#5}%
  \selectfont}%
\fi\endgroup%
{\renewcommand{\dashlinestretch}{30}
\begin{picture}(2508,123)(0,-10)
\thicklines
\dottedline{360}(54,54)(2454,54)
\end{picture}
}
}\hspace{-6pt}}
\newcommand{\PIdent} {\hspace{-6pt}\raisebox{-1pt}{\setlength{\unitlength}{0.00016667in}
\begingroup\makeatletter\ifx\SetFigFont\undefined%
\gdef\SetFigFont#1#2#3#4#5{%
  \reset@font\fontsize{#1}{#2pt}%
  \fontfamily{#3}\fontseries{#4}\fontshape{#5}%
  \selectfont}%
\fi\endgroup%
{\renewcommand{\dashlinestretch}{30}
\begin{picture}(4908,723)(0,-10)
\thicklines
\path(54,654)(4854,654)
\path(54,54)(4854,54)
\path(2154,54)(2754,54)
\blacken\path(2154.000,-96.000)(2754.000,54.000)(2154.000,204.000)(2154.000,-96.000)
\drawline(2754,654)(2754,654)
\path(2529,654)(1929,654)
\blacken\path(2529.000,804.000)(1929.000,654.000)(2529.000,504.000)(2529.000,804.000)
\end{picture}
}
}\hspace{-6pt}}
\newcommand{\PIdentII} {\hspace{-6pt}\raisebox{-1pt}{\setlength{\unitlength}{0.00016667in}
\begingroup\makeatletter\ifx\SetFigFont\undefined%
\gdef\SetFigFont#1#2#3#4#5{%
  \reset@font\fontsize{#1}{#2pt}%
  \fontfamily{#3}\fontseries{#4}\fontshape{#5}%
  \selectfont}%
\fi\endgroup%
{\renewcommand{\dashlinestretch}{30}
\begin{picture}(2508,723)(0,-10)
\thicklines
\path(54,654)(2454,654)
\path(54,54)(2454,54)
\end{picture}
}
}\hspace{-6pt}}
\newcommand{\PIdentAA} {\hspace{-6pt}\raisebox{-1pt}{\setlength{\unitlength}{0.00016667in}
\begingroup\makeatletter\ifx\SetFigFont\undefined%
\gdef\SetFigFont#1#2#3#4#5{%
  \reset@font\fontsize{#1}{#2pt}%
  \fontfamily{#3}\fontseries{#4}\fontshape{#5}%
  \selectfont}%
\fi\endgroup%
{\renewcommand{\dashlinestretch}{30}
\begin{picture}(2508,723)(0,-10)
\thicklines
\dashline{360.000}(54,654)(2454,654)
\dashline{360.000}(54,54)(2454,54)
\end{picture}
}
}\hspace{-6pt}}
\newcommand{\PIdentIA} {\hspace{-6pt}\raisebox{-1pt}{\setlength{\unitlength}{0.00016667in}
\begingroup\makeatletter\ifx\SetFigFont\undefined%
\gdef\SetFigFont#1#2#3#4#5{%
  \reset@font\fontsize{#1}{#2pt}%
  \fontfamily{#3}\fontseries{#4}\fontshape{#5}%
  \selectfont}%
\fi\endgroup%
{\renewcommand{\dashlinestretch}{30}
\begin{picture}(2508,723)(0,-10)
\thicklines
\path(54,654)(2454,654)
\dashline{360.000}(54,54)(2454,54)
\end{picture}
}
}\hspace{-6pt}}
\newcommand{\PIdentAB} {\hspace{-6pt}\raisebox{-1pt}{\setlength{\unitlength}{0.00016667in}
\begingroup\makeatletter\ifx\SetFigFont\undefined%
\gdef\SetFigFont#1#2#3#4#5{%
  \reset@font\fontsize{#1}{#2pt}%
  \fontfamily{#3}\fontseries{#4}\fontshape{#5}%
  \selectfont}%
\fi\endgroup%
{\renewcommand{\dashlinestretch}{30}
\begin{picture}(2508,723)(0,-10)
\thicklines
\dottedline{360}(54,54)(2454,54)
\dashline{360.000}(54,654)(2454,654)
\end{picture}
}
}\hspace{-6pt}}
\newcommand{\PTrace} {\hspace{-6pt}\raisebox{-1pt}{\setlength{\unitlength}{0.00016667in}
\begingroup\makeatletter\ifx\SetFigFont\undefined%
\gdef\SetFigFont#1#2#3#4#5{%
  \reset@font\fontsize{#1}{#2pt}%
  \fontfamily{#3}\fontseries{#4}\fontshape{#5}%
  \selectfont}%
\fi\endgroup%
{\renewcommand{\dashlinestretch}{30}
\begin{picture}(4908,723)(0,-10)
\thicklines
\path(54,654)(2154,654)(2154,54)(54,54)
\path(4854,654)(2754,654)(2754,54)(4854,54)
\path(3354,54)(4179,54)
\blacken\path(3579.000,-96.000)(4179.000,54.000)(3579.000,204.000)(3579.000,-96.000)
\path(4254,654)(3354,654)
\blacken\path(3954.000,804.000)(3354.000,654.000)(3954.000,504.000)(3954.000,804.000)
\path(654,54)(1554,54)
\blacken\path(954.000,-96.000)(1554.000,54.000)(954.000,204.000)(954.000,-96.000)
\path(1554,654)(729,654)
\blacken\path(1329.000,804.000)(729.000,654.000)(1329.000,504.000)(1329.000,804.000)
\end{picture}
}}\hspace{-6pt}}
\newcommand{\PTraceII} {\hspace{-6pt}\raisebox{-1pt}{\setlength{\unitlength}{0.00016667in}
\begingroup\makeatletter\ifx\SetFigFont\undefined%
\gdef\SetFigFont#1#2#3#4#5{%
  \reset@font\fontsize{#1}{#2pt}%
  \fontfamily{#3}\fontseries{#4}\fontshape{#5}%
  \selectfont}%
\fi\endgroup%
{\renewcommand{\dashlinestretch}{30}
\begin{picture}(2508,723)(0,-10)
\thicklines
\path(54,654)(954,654)(954,54)(54,54)
\path(2454,654)(1554,654)(1554,54)(2454,54)
\end{picture}
}
}\hspace{-6pt}}
\newcommand{\PTraceAA} {\hspace{-6pt}\raisebox{-1pt}{\setlength{\unitlength}{0.00016667in}
\begingroup\makeatletter\ifx\SetFigFont\undefined%
\gdef\SetFigFont#1#2#3#4#5{%
  \reset@font\fontsize{#1}{#2pt}%
  \fontfamily{#3}\fontseries{#4}\fontshape{#5}%
  \selectfont}%
\fi\endgroup%
{\renewcommand{\dashlinestretch}{30}
\begin{picture}(2508,723)(0,-10)
\thicklines
\dashline{360.000}(54,654)(954,654)(954,54)(54,54)
\dashline{360.000}(2454,654)(1554,654)(1554,54)(2454,54)
\end{picture}
}
}\hspace{-6pt}}
\newcommand{\PTraceSqr} {\hspace{-6pt}\raisebox{-1pt}{\setlength{\unitlength}{0.00016667in}
\begingroup\makeatletter\ifx\SetFigFont\undefined%
\gdef\SetFigFont#1#2#3#4#5{%
  \reset@font\fontsize{#1}{#2pt}%
  \fontfamily{#3}\fontseries{#4}\fontshape{#5}%
  \selectfont}%
\fi\endgroup%
{\renewcommand{\dashlinestretch}{30}
\begin{picture}(4908,723)(0,-10)
\thicklines
\path(2454,54)(2754,54)
\blacken\path(2154.000,-96.000)(2754.000,54.000)(2154.000,204.000)(2154.000,-96.000)
\path(2454,654)(2154,654)
\blacken\path(2754.000,804.000)(2154.000,654.000)(2754.000,504.000)(2754.000,804.000)
\path(654,54)(954,54)
\blacken\path(354.000,-96.000)(954.000,54.000)(354.000,204.000)(354.000,-96.000)
\path(354,654)(54,654)
\blacken\path(654.000,804.000)(54.000,654.000)(654.000,504.000)(654.000,804.000)
\path(4554,54)(4854,54)
\blacken\path(4254.000,-96.000)(4854.000,54.000)(4254.000,204.000)(4254.000,-96.000)
\path(4254,654)(3954,654)
\blacken\path(4554.000,804.000)(3954.000,654.000)(4554.000,504.000)(4554.000,804.000)
\path(54,654)(954,654)(954,54)(54,54)
\path(1554,654)(1554,54)(3354,54)
	(3354,654)(1554,654)
\path(4854,654)(3954,654)(3954,54)(4854,54)
\end{picture}
}}\hspace{-6pt}}
\newcommand{\PPhoton} {\hspace{-6pt}\raisebox{-1pt}{\setlength{\unitlength}{0.00016667in}
\begingroup\makeatletter\ifx\SetFigFont\undefined%
\gdef\SetFigFont#1#2#3#4#5{%
  \reset@font\fontsize{#1}{#2pt}%
  \fontfamily{#3}\fontseries{#4}\fontshape{#5}%
  \selectfont}%
\fi\endgroup%
{\renewcommand{\dashlinestretch}{30}
\begin{picture}(4908,735)(0,-10)
\thicklines
\path(54,360)(55,362)(58,368)
	(64,377)(71,390)(82,409)
	(96,431)(112,457)(130,486)
	(149,515)(169,544)(190,572)
	(211,597)(231,619)(251,638)
	(271,652)(291,661)(311,666)
	(332,665)(354,660)(370,653)
	(388,644)(406,632)(424,618)
	(444,600)(464,580)(484,557)
	(505,532)(527,504)(549,475)
	(571,444)(594,412)(617,379)
	(640,346)(663,313)(686,280)
	(709,249)(732,219)(755,190)
	(778,164)(801,140)(823,119)
	(846,101)(868,86)(889,75)
	(911,66)(933,62)(954,60)
	(975,62)(997,66)(1018,75)
	(1040,87)(1061,102)(1083,121)
	(1104,143)(1125,168)(1147,196)
	(1168,226)(1190,258)(1211,291)
	(1233,325)(1254,360)(1275,395)
	(1297,429)(1318,462)(1340,494)
	(1361,524)(1383,552)(1404,577)
	(1425,599)(1447,618)(1468,633)
	(1490,645)(1511,654)(1533,658)
	(1554,660)(1575,658)(1597,654)
	(1618,645)(1640,633)(1661,618)
	(1683,599)(1704,577)(1725,552)
	(1747,524)(1768,494)(1790,462)
	(1811,429)(1833,395)(1854,360)
	(1875,325)(1897,291)(1918,258)
	(1940,226)(1961,196)(1983,168)
	(2004,143)(2025,121)(2047,102)
	(2068,87)(2090,75)(2111,66)
	(2133,62)(2154,60)(2175,62)
	(2197,66)(2218,75)(2240,87)
	(2261,102)(2283,121)(2304,143)
	(2325,168)(2347,196)(2368,226)
	(2390,258)(2411,291)(2433,325)
	(2454,360)(2475,395)(2497,429)
	(2518,462)(2540,494)(2561,524)
	(2583,552)(2604,577)(2625,599)
	(2647,618)(2668,633)(2690,645)
	(2711,654)(2733,658)(2754,660)
	(2775,658)(2797,654)(2818,645)
	(2840,633)(2861,618)(2883,599)
	(2904,577)(2925,552)(2947,524)
	(2968,494)(2990,462)(3011,429)
	(3033,395)(3054,360)(3075,325)
	(3097,291)(3118,258)(3140,226)
	(3161,196)(3183,168)(3204,143)
	(3225,121)(3247,102)(3268,87)
	(3290,75)(3311,66)(3333,62)
	(3354,60)(3375,62)(3397,66)
	(3418,75)(3440,87)(3461,102)
	(3483,121)(3504,143)(3525,168)
	(3547,196)(3568,226)(3590,258)
	(3611,291)(3633,325)(3654,360)
	(3675,395)(3697,429)(3718,462)
	(3740,494)(3761,524)(3783,552)
	(3804,577)(3825,599)(3847,618)
	(3868,633)(3890,645)(3911,654)
	(3933,658)(3954,660)(3975,658)
	(3997,654)(4019,645)(4040,634)
	(4062,619)(4085,601)(4107,580)
	(4130,556)(4153,530)(4176,501)
	(4199,471)(4222,440)(4245,407)
	(4268,374)(4291,341)(4314,308)
	(4337,276)(4359,245)(4381,216)
	(4403,188)(4424,163)(4444,140)
	(4464,120)(4484,102)(4502,88)
	(4520,76)(4538,67)(4554,60)
	(4576,55)(4597,54)(4617,59)
	(4637,68)(4657,82)(4677,101)
	(4697,123)(4718,148)(4739,176)
	(4759,205)(4778,234)(4796,263)
	(4812,289)(4826,311)(4837,330)
	(4844,343)(4850,352)(4853,358)(4854,360)
\end{picture}
}}\hspace{-6pt}}

\newcommand{\VFund} {\hspace{-6pt}\raisebox{-12pt}{\setlength{\unitlength}{0.00016667in}
\begingroup\makeatletter\ifx\SetFigFont\undefined%
\gdef\SetFigFont#1#2#3#4#5{%
  \reset@font\fontsize{#1}{#2pt}%
  \fontfamily{#3}\fontseries{#4}\fontshape{#5}%
  \selectfont}%
\fi\endgroup%
{\renewcommand{\dashlinestretch}{30}
\begin{picture}(3708,2523)(0,-10)
\thicklines
\path(54,2454)(1554,2454)(1554,54)
\path(2154,54)(2154,2454)(3654,2454)
\path(654,2454)(1254,2454)
\path(654.000,2304.000)(1254.000,2454.000)(654.000,2604.000)
\path(2454,2454)(3054,2454)
\path(2454.000,2304.000)(3054.000,2454.000)(2454.000,2604.000)
\end{picture}
}
}\hspace{-6pt}}

\newcommand{\VAdjoint} {\hspace{-6pt}\raisebox{-12pt}{\setlength{\unitlength}{0.00016667in}
\begingroup\makeatletter\ifx\SetFigFont\undefined%
\gdef\SetFigFont#1#2#3#4#5{%
  \reset@font\fontsize{#1}{#2pt}%
  \fontfamily{#3}\fontseries{#4}\fontshape{#5}%
  \selectfont}%
\fi\endgroup%
{\renewcommand{\dashlinestretch}{30}
\begin{picture}(3708,2523)(0,-10)
\thicklines
\path(54,2454)(3654,2454)
\path(54,1854)(1554,1854)(1554,54)
\path(2154,54)(2154,1854)(3654,1854)
\blacken\path(2304.000,2604.000)(1704.000,2454.000)(2304.000,2304.000)(2304.000,2604.000)
\path(1704,2454)(2304,2454)
\blacken\path(654.000,1704.000)(1254.000,1854.000)(654.000,2004.000)(654.000,1704.000)
\path(1254,1854)(354,1854)
\path(2154,654)(2154,1554)
\blacken\path(2304.000,954.000)(2154.000,1554.000)(2004.000,954.000)(2304.000,954.000)
\end{picture}
}}\hspace{-6pt}}
\newcommand{\VAdjointX} {\hspace{-6pt}\raisebox{-12pt}{\setlength{\unitlength}{0.00016667in}
\begingroup\makeatletter\ifx\SetFigFont\undefined%
\gdef\SetFigFont#1#2#3#4#5{%
  \reset@font\fontsize{#1}{#2pt}%
  \fontfamily{#3}\fontseries{#4}\fontshape{#5}%
  \selectfont}%
\fi\endgroup%
{\renewcommand{\dashlinestretch}{30}
\begin{picture}(3708,2523)(0,-10)
\thicklines
\path(54,2454)(1554,2454)
\path(1554,1854)(1554,54)
\path(2154,2454)(3654,2454)
\path(2154,1854)(2154,54)
\path(654,1854)(1254,1854)
\blacken\path(654.000,1704.000)(1254.000,1854.000)(654.000,2004.000)(654.000,1704.000)
\path(3054,2454)(2454,2454)
\blacken\path(3054.000,2604.000)(2454.000,2454.000)(3054.000,2304.000)(3054.000,2604.000)
\path(2154,654)(2154,1254)
\blacken\path(2304.000,654.000)(2154.000,1254.000)(2004.000,654.000)(2304.000,654.000)
\path(54,1854)(3654,1854)
\path(1554,2454)(1556,2453)(1561,2451)
	(1570,2448)(1584,2442)(1602,2435)
	(1625,2426)(1651,2415)(1680,2404)
	(1711,2391)(1742,2377)(1773,2363)
	(1803,2350)(1831,2336)(1857,2322)
	(1881,2309)(1902,2296)(1921,2282)
	(1938,2269)(1954,2254)(1969,2238)
	(1982,2221)(1996,2202)(2009,2181)
	(2022,2157)(2036,2131)(2050,2103)
	(2063,2073)(2077,2042)(2091,2011)
	(2104,1980)(2115,1951)(2126,1925)
	(2135,1902)(2142,1884)(2148,1870)
	(2151,1861)(2153,1856)(2154,1854)
\path(2154,2454)(2152,2453)(2147,2451)
	(2138,2448)(2124,2442)(2106,2435)
	(2083,2426)(2057,2415)(2028,2404)
	(1997,2391)(1966,2377)(1935,2363)
	(1905,2350)(1877,2336)(1851,2322)
	(1827,2309)(1806,2296)(1787,2282)
	(1770,2269)(1754,2254)(1739,2238)
	(1726,2221)(1712,2202)(1699,2181)
	(1686,2157)(1672,2131)(1658,2103)
	(1645,2073)(1631,2042)(1617,2011)
	(1604,1980)(1593,1951)(1582,1925)
	(1573,1902)(1566,1884)(1560,1870)
	(1557,1861)(1555,1856)(1554,1854)
\end{picture}
}}\hspace{-6pt}}
\newcommand{\VQuark} {\hspace{-6pt}\raisebox{-12pt}{\setlength{\unitlength}{0.00016667in}
\begingroup\makeatletter\ifx\SetFigFont\undefined%
\gdef\SetFigFont#1#2#3#4#5{%
  \reset@font\fontsize{#1}{#2pt}%
  \fontfamily{#3}\fontseries{#4}\fontshape{#5}%
  \selectfont}%
\fi\endgroup%
{\renewcommand{\dashlinestretch}{30}
\begin{picture}(3708,2523)(0,-10)
\thicklines
\path(54,2454)(2154,2454)
\path(1554.000,2304.000)(2154.000,2454.000)(1554.000,2604.000)
\path(1854,2454)(3654,2454)
\path(1854,2454)(1856,2453)(1862,2450)
	(1871,2444)(1884,2437)(1903,2426)
	(1925,2412)(1951,2396)(1980,2378)
	(2009,2359)(2038,2339)(2066,2318)
	(2091,2297)(2113,2277)(2132,2257)
	(2146,2237)(2155,2217)(2160,2197)
	(2159,2176)(2154,2154)(2147,2138)
	(2138,2120)(2126,2102)(2112,2084)
	(2094,2064)(2074,2044)(2051,2024)
	(2026,2003)(1998,1981)(1969,1959)
	(1938,1937)(1906,1914)(1873,1891)
	(1840,1868)(1807,1845)(1774,1822)
	(1743,1799)(1713,1776)(1684,1753)
	(1658,1730)(1634,1707)(1613,1685)
	(1595,1662)(1580,1640)(1569,1619)
	(1560,1597)(1556,1575)(1554,1554)
	(1556,1533)(1560,1511)(1569,1490)
	(1581,1468)(1596,1447)(1615,1425)
	(1637,1404)(1662,1383)(1690,1361)
	(1720,1340)(1752,1318)(1785,1297)
	(1819,1275)(1854,1254)(1889,1233)
	(1923,1211)(1956,1190)(1988,1168)
	(2018,1147)(2046,1125)(2071,1104)
	(2093,1083)(2112,1061)(2127,1040)
	(2139,1018)(2148,997)(2152,975)
	(2154,954)(2152,933)(2148,911)
	(2139,889)(2128,868)(2113,846)
	(2095,823)(2074,801)(2050,778)
	(2024,755)(1995,732)(1965,709)
	(1934,686)(1901,663)(1868,640)
	(1835,617)(1802,594)(1770,571)
	(1739,549)(1710,527)(1682,505)
	(1657,484)(1634,464)(1614,444)
	(1596,424)(1582,406)(1570,388)
	(1561,370)(1554,354)(1549,332)
	(1548,311)(1553,291)(1562,271)
	(1576,251)(1595,231)(1617,211)
	(1642,190)(1670,169)(1699,149)
	(1728,130)(1757,112)(1783,96)
	(1805,82)(1824,71)(1837,64)
	(1846,58)(1852,55)(1854,54)
\end{picture}
}}\hspace{-6pt}}
\newcommand{\VPhoton} {\hspace{-6pt}\raisebox{-12pt}{\setlength{\unitlength}{0.00016667in}
\begingroup\makeatletter\ifx\SetFigFont\undefined%
\gdef\SetFigFont#1#2#3#4#5{%
  \reset@font\fontsize{#1}{#2pt}%
  \fontfamily{#3}\fontseries{#4}\fontshape{#5}%
  \selectfont}%
\fi\endgroup%
{\renewcommand{\dashlinestretch}{30}
\begin{picture}(4308,2529)(0,-10)
\thicklines
\put(2154,1854){\blacken\ellipse{306}{306}}
\put(2154,1854){\ellipse{306}{306}}
\path(54,2154)(55,2156)(58,2162)
	(64,2171)(71,2184)(82,2203)
	(96,2225)(112,2251)(130,2280)
	(149,2309)(169,2338)(190,2366)
	(211,2391)(231,2413)(251,2432)
	(271,2446)(291,2455)(311,2460)
	(332,2459)(354,2454)(370,2447)
	(388,2438)(406,2426)(424,2412)
	(444,2394)(464,2374)(484,2351)
	(505,2326)(527,2298)(549,2269)
	(571,2238)(594,2206)(617,2173)
	(640,2140)(663,2107)(686,2074)
	(709,2043)(732,2013)(755,1984)
	(778,1958)(801,1934)(823,1913)
	(846,1895)(868,1880)(889,1869)
	(911,1860)(933,1856)(954,1854)
	(975,1856)(997,1860)(1018,1869)
	(1040,1881)(1061,1896)(1083,1915)
	(1104,1937)(1125,1962)(1147,1990)
	(1168,2020)(1190,2052)(1211,2085)
	(1233,2119)(1254,2154)(1275,2189)
	(1297,2223)(1318,2256)(1340,2288)
	(1361,2318)(1383,2346)(1404,2371)
	(1425,2393)(1447,2412)(1468,2427)
	(1490,2439)(1511,2448)(1533,2452)
	(1554,2454)(1575,2452)(1597,2448)
	(1618,2439)(1640,2427)(1661,2412)
	(1683,2393)(1704,2371)(1725,2346)
	(1747,2318)(1768,2288)(1790,2256)
	(1811,2223)(1833,2189)(1854,2154)
	(1875,2119)(1897,2085)(1918,2052)
	(1940,2020)(1961,1990)(1983,1962)
	(2004,1937)(2025,1915)(2047,1896)
	(2068,1881)(2090,1869)(2111,1860)
	(2133,1856)(2154,1854)(2175,1856)
	(2197,1860)(2218,1869)(2240,1881)
	(2261,1896)(2283,1915)(2304,1937)
	(2325,1962)(2347,1990)(2368,2020)
	(2390,2052)(2411,2085)(2433,2119)
	(2454,2154)(2475,2189)(2497,2223)
	(2518,2256)(2540,2288)(2561,2318)
	(2583,2346)(2604,2371)(2625,2393)
	(2647,2412)(2668,2427)(2690,2439)
	(2711,2448)(2733,2452)(2754,2454)
	(2775,2452)(2797,2448)(2818,2439)
	(2840,2427)(2861,2412)(2883,2393)
	(2904,2371)(2925,2346)(2947,2318)
	(2968,2288)(2990,2256)(3011,2223)
	(3033,2189)(3054,2154)(3075,2119)
	(3097,2085)(3118,2052)(3140,2020)
	(3161,1990)(3183,1962)(3204,1937)
	(3225,1915)(3247,1896)(3268,1881)
	(3290,1869)(3311,1860)(3333,1856)
	(3354,1854)(3375,1856)(3397,1860)
	(3419,1869)(3440,1880)(3462,1895)
	(3485,1913)(3507,1934)(3530,1958)
	(3553,1984)(3576,2013)(3599,2043)
	(3622,2074)(3645,2107)(3668,2140)
	(3691,2173)(3714,2206)(3737,2238)
	(3759,2269)(3781,2298)(3803,2326)
	(3824,2351)(3844,2374)(3864,2394)
	(3884,2412)(3902,2426)(3920,2438)
	(3938,2447)(3954,2454)(3976,2459)
	(3997,2460)(4017,2455)(4037,2446)
	(4057,2432)(4077,2413)(4097,2391)
	(4118,2366)(4139,2338)(4159,2309)
	(4178,2280)(4196,2251)(4212,2225)
	(4226,2203)(4237,2184)(4244,2171)
	(4250,2162)(4253,2156)(4254,2154)
\path(2154,1854)(2152,1853)(2146,1850)
	(2137,1844)(2124,1837)(2105,1826)
	(2083,1812)(2057,1796)(2028,1778)
	(1999,1759)(1970,1739)(1942,1718)
	(1917,1697)(1895,1677)(1876,1657)
	(1862,1637)(1853,1617)(1848,1597)
	(1849,1576)(1854,1554)(1861,1538)
	(1870,1520)(1882,1502)(1896,1484)
	(1914,1464)(1934,1444)(1957,1424)
	(1982,1403)(2010,1381)(2039,1359)
	(2070,1337)(2102,1314)(2135,1291)
	(2168,1268)(2201,1245)(2234,1222)
	(2265,1199)(2295,1176)(2324,1153)
	(2350,1130)(2374,1107)(2395,1085)
	(2413,1062)(2428,1040)(2439,1019)
	(2448,997)(2452,975)(2454,954)
	(2452,933)(2448,911)(2439,889)
	(2428,868)(2413,846)(2395,823)
	(2374,801)(2350,778)(2324,755)
	(2295,732)(2265,709)(2234,686)
	(2201,663)(2168,640)(2135,617)
	(2102,594)(2070,571)(2039,549)
	(2010,527)(1982,505)(1957,484)
	(1934,464)(1914,444)(1896,424)
	(1882,406)(1870,388)(1861,370)
	(1854,354)(1849,332)(1848,311)
	(1853,291)(1862,271)(1876,251)
	(1895,231)(1917,211)(1942,190)
	(1970,169)(1999,149)(2028,130)
	(2057,112)(2083,96)(2105,82)
	(2124,71)(2137,64)(2146,58)
	(2152,55)(2154,54)
\end{picture}
}}\hspace{-6pt}}
\newcommand{\VBifund} {\hspace{-6pt}\raisebox{-12pt}{\setlength{\unitlength}{0.00016667in}
\begingroup\makeatletter\ifx\SetFigFont\undefined%
\gdef\SetFigFont#1#2#3#4#5{%
  \reset@font\fontsize{#1}{#2pt}%
  \fontfamily{#3}\fontseries{#4}\fontshape{#5}%
  \selectfont}%
\fi\endgroup%
{\renewcommand{\dashlinestretch}{30}
\begin{picture}(3708,2523)(0,-10)
\thicklines
\path(54,1854)(1554,1854)(1554,54)
\path(2154,54)(2154,1854)(3654,1854)
\blacken\path(2304.000,2604.000)(1704.000,2454.000)(2304.000,2304.000)(2304.000,2604.000)
\path(1704,2454)(2304,2454)
\blacken\path(654.000,1704.000)(1254.000,1854.000)(654.000,2004.000)(654.000,1704.000)
\path(1254,1854)(354,1854)
\path(2154,654)(2154,1554)
\blacken\path(2304.000,954.000)(2154.000,1554.000)(2004.000,954.000)(2304.000,954.000)
\dashline{360.000}(54,2454)(3654,2454)
\end{picture}
}}\hspace{-6pt}}
\newcommand{\VBifundX} {\hspace{-6pt}\raisebox{-12pt}{\setlength{\unitlength}{0.00016667in}
\begingroup\makeatletter\ifx\SetFigFont\undefined%
\gdef\SetFigFont#1#2#3#4#5{%
  \reset@font\fontsize{#1}{#2pt}%
  \fontfamily{#3}\fontseries{#4}\fontshape{#5}%
  \selectfont}%
\fi\endgroup%
{\renewcommand{\dashlinestretch}{30}
\begin{picture}(3708,2523)(0,-10)
\thicklines
\path(54,2454)(1554,2454)
\path(1554,1854)(1554,54)
\path(2154,2454)(3654,2454)
\path(2154,1854)(2154,54)
\path(654,1854)(1254,1854)
\blacken\path(654.000,1704.000)(1254.000,1854.000)(654.000,2004.000)(654.000,1704.000)
\path(3054,2454)(2454,2454)
\blacken\path(3054.000,2604.000)(2454.000,2454.000)(3054.000,2304.000)(3054.000,2604.000)
\path(2154,654)(2154,1254)
\blacken\path(2304.000,654.000)(2154.000,1254.000)(2004.000,654.000)(2304.000,654.000)
\dottedline{360}(54,1854)(3654,1854)
\path(1554,2454)(1556,2453)(1561,2451)
	(1570,2448)(1584,2442)(1602,2435)
	(1625,2426)(1651,2415)(1680,2404)
	(1711,2391)(1742,2377)(1773,2363)
	(1803,2350)(1831,2336)(1857,2322)
	(1881,2309)(1902,2296)(1921,2282)
	(1938,2269)(1954,2254)(1969,2238)
	(1982,2221)(1996,2202)(2009,2181)
	(2022,2157)(2036,2131)(2050,2103)
	(2063,2073)(2077,2042)(2091,2011)
	(2104,1980)(2115,1951)(2126,1925)
	(2135,1902)(2142,1884)(2148,1870)
	(2151,1861)(2153,1856)(2154,1854)
\path(2154,2454)(2152,2453)(2147,2451)
	(2138,2448)(2124,2442)(2106,2435)
	(2083,2426)(2057,2415)(2028,2404)
	(1997,2391)(1966,2377)(1935,2363)
	(1905,2350)(1877,2336)(1851,2322)
	(1827,2309)(1806,2296)(1787,2282)
	(1770,2269)(1754,2254)(1739,2238)
	(1726,2221)(1712,2202)(1699,2181)
	(1686,2157)(1672,2131)(1658,2103)
	(1645,2073)(1631,2042)(1617,2011)
	(1604,1980)(1593,1951)(1582,1925)
	(1573,1902)(1566,1884)(1560,1870)
	(1557,1861)(1555,1856)(1554,1854)
\end{picture}
}}\hspace{-6pt}}

\newcommand{\sChannelFund} {\hspace{-6pt}\raisebox{-16pt}{\setlength{\unitlength}{0.00016667in}
\begingroup\makeatletter\ifx\SetFigFont\undefined%
\gdef\SetFigFont#1#2#3#4#5{%
  \reset@font\fontsize{#1}{#2pt}%
  \fontfamily{#3}\fontseries{#4}\fontshape{#5}%
  \selectfont}%
\fi\endgroup%
{\renewcommand{\dashlinestretch}{30}
\begin{picture}(3708,2523)(0,-10)
\thicklines
\path(54,54)(3654,54)
\path(54,654)(1554,654)(1554,2454)
\path(2154,2454)(2154,654)(3654,654)
\path(54,2454)(1554,2454)
\path(2154,2454)(3654,2454)
\end{picture}
}
}\hspace{-6pt}}
\newcommand{\sChannelFundX} {\hspace{-6pt}\raisebox{-16pt}{\setlength{\unitlength}{0.00016667in}
\begingroup\makeatletter\ifx\SetFigFont\undefined%
\gdef\SetFigFont#1#2#3#4#5{%
  \reset@font\fontsize{#1}{#2pt}%
  \fontfamily{#3}\fontseries{#4}\fontshape{#5}%
  \selectfont}%
\fi\endgroup%
{\renewcommand{\dashlinestretch}{30}
\begin{picture}(3708,2523)(0,-10)
\thicklines
\path(54,54)(1554,54)
\path(2154,54)(3654,54)
\path(2154,654)(2154,2454)
\path(2154,2454)(3654,2454)
\path(54,2454)(1554,2454)(1554,654)
\path(3654,654)(54,654)
\path(1554,54)(1556,55)(1561,57)
	(1570,60)(1584,66)(1602,73)
	(1625,82)(1651,93)(1680,104)
	(1711,117)(1742,131)(1773,145)
	(1803,158)(1831,172)(1857,186)
	(1881,199)(1902,212)(1921,226)
	(1938,239)(1954,254)(1969,270)
	(1982,287)(1996,306)(2009,327)
	(2022,351)(2036,377)(2050,405)
	(2063,435)(2077,466)(2091,497)
	(2104,528)(2115,557)(2126,583)
	(2135,606)(2142,624)(2148,638)
	(2151,647)(2153,652)(2154,654)
\path(2154,54)(2152,55)(2147,57)
	(2138,60)(2124,66)(2106,73)
	(2083,82)(2057,93)(2028,104)
	(1997,117)(1966,131)(1935,145)
	(1905,158)(1877,172)(1851,186)
	(1827,199)(1806,212)(1787,226)
	(1770,239)(1754,254)(1739,270)
	(1726,287)(1712,306)(1699,327)
	(1686,351)(1672,377)(1658,405)
	(1645,435)(1631,466)(1617,497)
	(1604,528)(1593,557)(1582,583)
	(1573,606)(1566,624)(1560,638)
	(1557,647)(1555,652)(1554,654)
\end{picture}
}}\hspace{-6pt}}
\newcommand{\tChannelFund} {\hspace{-6pt}\raisebox{-16pt}{\setlength{\unitlength}{0.00016667in}
\begingroup\makeatletter\ifx\SetFigFont\undefined%
\gdef\SetFigFont#1#2#3#4#5{%
  \reset@font\fontsize{#1}{#2pt}%
  \fontfamily{#3}\fontseries{#4}\fontshape{#5}%
  \selectfont}%
\fi\endgroup%
{\renewcommand{\dashlinestretch}{30}
\begin{picture}(3708,2507)(0,-10)
\thicklines
\put(-96.000,1688.000){\arc{2121.320}{0.1419}{1.4289}}
\put(3691.500,1575.500){\arc{3075.914}{1.5952}{3.1172}}
\put(3804.000,1688.000){\arc{2121.320}{1.7127}{2.9997}}
\put(16.500,1575.500){\arc{3075.914}{0.0244}{1.5464}}
\path(2154,1538)(2154,2438)(1554,2438)(1554,1538)
\path(954,1538)(954,2438)(54,2438)
\path(2754,1538)(2754,2438)(3654,2438)
\end{picture}
}
}\hspace{-6pt}}
\newcommand{\uChannelFund} {\hspace{-6pt}\raisebox{-16pt}{\setlength{\unitlength}{0.00016667in}
\begingroup\makeatletter\ifx\SetFigFont\undefined%
\gdef\SetFigFont#1#2#3#4#5{%
  \reset@font\fontsize{#1}{#2pt}%
  \fontfamily{#3}\fontseries{#4}\fontshape{#5}%
  \selectfont}%
\fi\endgroup%
{\renewcommand{\dashlinestretch}{30}
\begin{picture}(3708,2523)(0,-10)
\thicklines
\put(2841.500,1941.500){\arc{2580.940}{1.6387}{3.0737}}
\put(866.500,1941.500){\arc{2580.940}{0.0679}{1.5029}}
\put(654.000,2154.000){\arc{4242.641}{0.1419}{1.4289}}
\put(3054.000,2154.000){\arc{4242.641}{1.7127}{2.9997}}
\path(54,2454)(954,2454)(954,1854)
\path(1554,1854)(1554,2454)(2154,2454)(2154,1854)
\path(3654,2454)(2754,2454)(2754,1854)
\path(2754,54)(3654,54)
\path(2754,654)(3654,654)
\path(954,54)(54,54)
\path(954,654)(54,654)
\end{picture}
}}\hspace{-6pt}}
\newcommand{\sChannelPhoton} {\epsFig{-16pt}{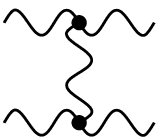}}
\newcommand{\tChannelPhoton} {\epsFig{-16pt}{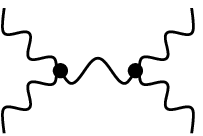}}
\newcommand{\uChannelPhoton} {\epsFig{-16pt}{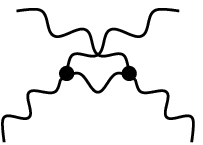}}

\newcommand{\PTraceVAdjoint} {\hspace{-6pt}\raisebox{-12pt}{\setlength{\unitlength}{0.00016667in}
\begingroup\makeatletter\ifx\SetFigFont\undefined%
\gdef\SetFigFont#1#2#3#4#5{%
  \reset@font\fontsize{#1}{#2pt}%
  \fontfamily{#3}\fontseries{#4}\fontshape{#5}%
  \selectfont}%
\fi\endgroup%
{\renewcommand{\dashlinestretch}{30}
\begin{picture}(3108,2523)(0,-10)
\thicklines
\path(54,1554)(954,1554)(954,954)(54,954)
\path(2454,2454)(2454,1554)(1554,1554)
	(1554,954)(2454,954)(2454,54)
\path(3054,2454)(3054,54)
\end{picture}
}
}\hspace{-6pt}}
\newcommand{\PTraceVAdjointX} {\hspace{-6pt}\raisebox{-12pt}{\setlength{\unitlength}{0.00016667in}
\begingroup\makeatletter\ifx\SetFigFont\undefined%
\gdef\SetFigFont#1#2#3#4#5{%
  \reset@font\fontsize{#1}{#2pt}%
  \fontfamily{#3}\fontseries{#4}\fontshape{#5}%
  \selectfont}%
\fi\endgroup%
{\renewcommand{\dashlinestretch}{30}
\begin{picture}(3108,2523)(0,-10)
\thicklines
\path(3054,2454)(3054,1554)
\path(2454,1554)(1554,1554)
\path(3054,954)(3054,54)
\path(2454,954)(1554,954)
\path(2454,2454)(2454,54)
\drawline(1554,1554)(1554,1554)
\path(1554,1554)(1554,954)
\path(54,1554)(954,1554)(954,954)(54,954)
\path(3054,1554)(3053,1552)(3051,1547)
	(3048,1538)(3042,1524)(3035,1506)
	(3026,1483)(3015,1457)(3004,1428)
	(2991,1397)(2977,1366)(2963,1335)
	(2950,1305)(2936,1277)(2922,1251)
	(2909,1227)(2896,1206)(2882,1187)
	(2869,1170)(2854,1154)(2838,1139)
	(2821,1126)(2802,1112)(2781,1099)
	(2757,1086)(2731,1072)(2703,1058)
	(2673,1045)(2642,1031)(2611,1017)
	(2580,1004)(2551,993)(2525,982)
	(2502,973)(2484,966)(2470,960)
	(2461,957)(2456,955)(2454,954)
\path(3054,954)(3053,956)(3051,961)
	(3048,970)(3042,984)(3035,1002)
	(3026,1025)(3015,1051)(3004,1080)
	(2991,1111)(2977,1142)(2963,1173)
	(2950,1203)(2936,1231)(2922,1257)
	(2909,1281)(2896,1302)(2882,1321)
	(2869,1338)(2854,1354)(2838,1369)
	(2821,1382)(2802,1396)(2781,1409)
	(2757,1422)(2731,1436)(2703,1450)
	(2673,1463)(2642,1477)(2611,1491)
	(2580,1504)(2551,1515)(2525,1526)
	(2502,1535)(2484,1542)(2470,1548)
	(2461,1551)(2456,1553)(2454,1554)
\end{picture}
}}\hspace{-6pt}}
\newcommand{\PTraceVBifund} {\hspace{-6pt}\raisebox{-12pt}{\setlength{\unitlength}{0.00016667in}
\begingroup\makeatletter\ifx\SetFigFont\undefined%
\gdef\SetFigFont#1#2#3#4#5{%
  \reset@font\fontsize{#1}{#2pt}%
  \fontfamily{#3}\fontseries{#4}\fontshape{#5}%
  \selectfont}%
\fi\endgroup%
{\renewcommand{\dashlinestretch}{30}
\begin{picture}(3108,2523)(0,-10)
\thicklines
\path(54,1554)(954,1554)(954,954)(54,954)
\path(2454,2454)(2454,1554)(1554,1554)
	(1554,954)(2454,954)(2454,54)
\dashline{360.000}(3054,2454)(3054,54)
\end{picture}
}
}\hspace{-6pt}}
\newcommand{\PTraceVBifundX} {\hspace{-6pt}\raisebox{-12pt}{\setlength{\unitlength}{0.00016667in}
\begingroup\makeatletter\ifx\SetFigFont\undefined%
\gdef\SetFigFont#1#2#3#4#5{%
  \reset@font\fontsize{#1}{#2pt}%
  \fontfamily{#3}\fontseries{#4}\fontshape{#5}%
  \selectfont}%
\fi\endgroup%
{\renewcommand{\dashlinestretch}{30}
\begin{picture}(3108,2523)(0,-10)
\thicklines
\path(3054,2454)(3054,1554)
\path(2454,1554)(1554,1554)
\path(3054,954)(3054,54)
\path(2454,954)(1554,954)
\drawline(1554,1554)(1554,1554)
\path(1554,1554)(1554,954)
\path(54,1554)(954,1554)(954,954)(54,954)
\dottedline{360}(2454,2454)(2454,54)
\path(3054,1554)(3053,1552)(3051,1547)
	(3048,1538)(3042,1524)(3035,1506)
	(3026,1483)(3015,1457)(3004,1428)
	(2991,1397)(2977,1366)(2963,1335)
	(2950,1305)(2936,1277)(2922,1251)
	(2909,1227)(2896,1206)(2882,1187)
	(2869,1170)(2854,1154)(2838,1139)
	(2821,1126)(2802,1112)(2781,1099)
	(2757,1086)(2731,1072)(2703,1058)
	(2673,1045)(2642,1031)(2611,1017)
	(2580,1004)(2551,993)(2525,982)
	(2502,973)(2484,966)(2470,960)
	(2461,957)(2456,955)(2454,954)
\path(3054,954)(3053,956)(3051,961)
	(3048,970)(3042,984)(3035,1002)
	(3026,1025)(3015,1051)(3004,1080)
	(2991,1111)(2977,1142)(2963,1173)
	(2950,1203)(2936,1231)(2922,1257)
	(2909,1281)(2896,1302)(2882,1321)
	(2869,1338)(2854,1354)(2838,1369)
	(2821,1382)(2802,1396)(2781,1409)
	(2757,1422)(2731,1436)(2703,1450)
	(2673,1463)(2642,1477)(2611,1491)
	(2580,1504)(2551,1515)(2525,1526)
	(2502,1535)(2484,1542)(2470,1548)
	(2461,1551)(2456,1553)(2454,1554)
\end{picture}
}}\hspace{-6pt}}

\newcommand{\POneLoopFund} {\hspace{-6pt}\raisebox{-8pt}{\setlength{\unitlength}{0.00016667in}
\begingroup\makeatletter\ifx\SetFigFont\undefined%
\gdef\SetFigFont#1#2#3#4#5{%
  \reset@font\fontsize{#1}{#2pt}%
  \fontfamily{#3}\fontseries{#4}\fontshape{#5}%
  \selectfont}%
\fi\endgroup%
{\renewcommand{\dashlinestretch}{30}
\begin{picture}(3708,1893)(0,-10)
\thicklines
\path(1854,1764)(1554,1764)
\blacken\path(2154.000,1914.000)(1554.000,1764.000)(2154.000,1614.000)(2154.000,1914.000)
\path(1854,114)(2154,114)
\blacken\path(1554.000,-36.000)(2154.000,114.000)(1554.000,264.000)(1554.000,-36.000)
\put(1854.000,1070.250){\arc{1537.500}{3.3629}{6.0619}}
\put(1854.000,807.750){\arc{1537.500}{0.2213}{2.9203}}
\path(54,1239)(1104,1239)
\path(54,639)(1104,639)
\path(2604,639)(3654,639)
\path(2604,1239)(3654,1239)
\end{picture}
}}\hspace{-6pt}}
\newcommand{\POneLoopOO} {\hspace{-6pt}\raisebox{-16pt}{\setlength{\unitlength}{0.00016667in}
\begingroup\makeatletter\ifx\SetFigFont\undefined%
\gdef\SetFigFont#1#2#3#4#5{%
  \reset@font\fontsize{#1}{#2pt}%
  \fontfamily{#3}\fontseries{#4}\fontshape{#5}%
  \selectfont}%
\fi\endgroup%
{\renewcommand{\dashlinestretch}{30}
\begin{picture}(4908,3092)(0,-10)
\thicklines
\put(2454.000,1591.906){\arc{2892.188}{3.3126}{6.1122}}
\put(2454.000,1484.094){\arc{2892.188}{0.1710}{2.9706}}
\put(2454,1538){\ellipse{1800}{1800}}
\path(54,1838)(1029,1838)
\path(54,1238)(1029,1238)
\path(3879,1838)(4854,1838)
\path(3879,1238)(4854,1238)
\end{picture}
}
}\hspace{-6pt}}
\newcommand{\POneLoopXO} {\hspace{-6pt}\raisebox{-16pt}{\setlength{\unitlength}{0.00016667in}
\begingroup\makeatletter\ifx\SetFigFont\undefined%
\gdef\SetFigFont#1#2#3#4#5{%
  \reset@font\fontsize{#1}{#2pt}%
  \fontfamily{#3}\fontseries{#4}\fontshape{#5}%
  \selectfont}%
\fi\endgroup%
{\renewcommand{\dashlinestretch}{30}
\begin{picture}(4908,3048)(0,-10)
\thicklines
\put(2465.413,1516.000){\arc{1777.174}{3.4860}{9.0804}}
\put(2431.885,1516.000){\arc{2955.768}{0.2044}{6.0788}}
\put(991.500,1178.500){\arc{1277.204}{4.7711}{6.2244}}
\put(991.500,1853.500){\arc{1277.204}{0.0588}{1.5120}}
\path(3879,1816)(4854,1816)
\path(3879,1216)(4854,1216)
\path(54,1816)(1029,1816)
\path(54,1216)(1029,1216)
\end{picture}
}
}\hspace{-6pt}}
\newcommand{\POneLoopOX} {\hspace{-6pt}\raisebox{-16pt}{\setlength{\unitlength}{0.00016667in}
\begingroup\makeatletter\ifx\SetFigFont\undefined%
\gdef\SetFigFont#1#2#3#4#5{%
  \reset@font\fontsize{#1}{#2pt}%
  \fontfamily{#3}\fontseries{#4}\fontshape{#5}%
  \selectfont}%
\fi\endgroup%
{\renewcommand{\dashlinestretch}{30}
\begin{picture}(4908,3048)(0,-10)
\thicklines
\put(2442.587,1516.000){\arc{1777.174}{0.3444}{5.9388}}
\put(2476.115,1516.000){\arc{2955.768}{3.3460}{9.2204}}
\put(3916.500,1178.500){\arc{1277.204}{3.2003}{4.6536}}
\put(3916.500,1853.500){\arc{1277.204}{1.6296}{3.0828}}
\path(1029,1816)(54,1816)
\path(1029,1216)(54,1216)
\path(4854,1816)(3879,1816)
\path(4854,1216)(3879,1216)
\end{picture}
}
}\hspace{-6pt}}
\newcommand{\POneLoopXX} {\hspace{-6pt}\raisebox{-16pt}{\setlength{\unitlength}{0.00016667in}
\begingroup\makeatletter\ifx\SetFigFont\undefined%
\gdef\SetFigFont#1#2#3#4#5{%
  \reset@font\fontsize{#1}{#2pt}%
  \fontfamily{#3}\fontseries{#4}\fontshape{#5}%
  \selectfont}%
\fi\endgroup%
{\renewcommand{\dashlinestretch}{30}
\begin{picture}(4908,3089)(0,-10)
\thicklines
\put(3916.500,1199.500){\arc{1277.204}{3.2003}{4.6536}}
\put(3916.500,1874.500){\arc{1277.204}{1.6296}{3.0828}}
\put(2454.000,1569.812){\arc{1734.375}{3.4548}{5.9700}}
\put(2454.000,1504.188){\arc{1734.375}{0.3132}{2.8284}}
\put(991.500,1199.500){\arc{1277.204}{4.7711}{6.2244}}
\put(991.500,1874.500){\arc{1277.204}{0.0588}{1.6296}}
\put(2454,1537){\ellipse{3000}{3000}}
\path(1029,1837)(54,1837)
\path(1029,1237)(54,1237)
\path(4854,1837)(3879,1837)
\path(4854,1237)(3879,1237)
\end{picture}
}}\hspace{-6pt}}
\newcommand{\POneLoopBifund} {\epsFig{-16pt}{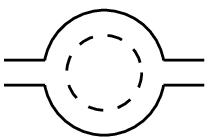}}
\newcommand{\POneLoopBifundTrace} {\epsFig{-16pt}{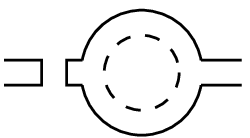}}
\newcommand{\PBifundOneLoop} {\hspace{-6pt}\raisebox{-2pt}{\setlength{\unitlength}{0.00016667in}
\begingroup\makeatletter\ifx\SetFigFont\undefined%
\gdef\SetFigFont#1#2#3#4#5{%
  \reset@font\fontsize{#1}{#2pt}%
  \fontfamily{#3}\fontseries{#4}\fontshape{#5}%
  \selectfont}%
\fi\endgroup%
{\renewcommand{\dashlinestretch}{30}
\begin{picture}(4908,2208)(0,-10)
\thicklines
\put(2454.000,654.000){\arc{3000.000}{3.1416}{6.2832}}
\put(2454.000,654.000){\arc{1800.000}{3.1416}{6.2832}}
\dashline{360.000}(54,54)(4854,54)
\path(54,654)(954,654)
\path(1554,654)(3354,654)
\path(3954,654)(4854,654)
\end{picture}
}
}\hspace{-6pt}}
\newcommand{\PBifundOneLoopTrace} {\hspace{-6pt}\raisebox{-2pt}{\setlength{\unitlength}{0.00016667in}
\begingroup\makeatletter\ifx\SetFigFont\undefined%
\gdef\SetFigFont#1#2#3#4#5{%
  \reset@font\fontsize{#1}{#2pt}%
  \fontfamily{#3}\fontseries{#4}\fontshape{#5}%
  \selectfont}%
\fi\endgroup%
{\renewcommand{\dashlinestretch}{30}
\begin{picture}(4908,2223)(0,-10)
\thicklines
\put(2518.286,632.572){\arc{3128.864}{3.1553}{4.4774}}
\put(2389.714,632.572){\arc{3128.866}{4.9474}{6.2695}}
\put(2529.000,754.000){\arc{1662.077}{4.9866}{6.4038}}
\put(2379.000,754.000){\arc{1662.077}{3.0210}{4.4382}}
\dashline{360.000}(54,54)(4854,54)
\path(54,654)(954,654)
\path(1554,654)(3354,654)
\path(3954,654)(4854,654)
\path(2154,2154)(2154,1554)
\path(2754,2154)(2754,1554)
\end{picture}
}}\hspace{-6pt}}
\newcommand{\PBifundOneLoopYukawa} {\epsFig{-16pt}{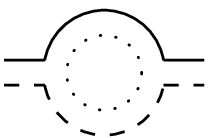}}

\newcommand{\VOneLoopOOO} {\hspace{-6pt}\raisebox{-24pt}{\setlength{\unitlength}{0.00016667in}
\begingroup\makeatletter\ifx\SetFigFont\undefined%
\gdef\SetFigFont#1#2#3#4#5{%
  \reset@font\fontsize{#1}{#2pt}%
  \fontfamily{#3}\fontseries{#4}\fontshape{#5}%
  \selectfont}%
\fi\endgroup%
{\renewcommand{\dashlinestretch}{30}
\begin{picture}(4608,4623)(0,-10)
\thicklines
\put(2304.000,2219.625){\arc{3468.750}{4.1154}{5.3094}}
\put(2184.000,2520.000){\arc{3242.688}{5.8477}{7.5919}}
\put(2424.000,2520.000){\arc{3242.688}{1.8328}{3.5771}}
\put(2304,2454){\ellipse{1800}{1800}}
\path(2004,54)(2004,954)
\path(2604,54)(2604,954)
\path(3279,3654)(4179,4554)
\path(3654,3204)(4554,4104)
\path(1329,3654)(429,4554)
\path(954,3204)(54,4104)
\end{picture}
}}\hspace{-6pt}}
\newcommand{\VOneLoopOXO} {\hspace{-6pt}\raisebox{-24pt}{\setlength{\unitlength}{0.00016667in}
\begingroup\makeatletter\ifx\SetFigFont\undefined%
\gdef\SetFigFont#1#2#3#4#5{%
  \reset@font\fontsize{#1}{#2pt}%
  \fontfamily{#3}\fontseries{#4}\fontshape{#5}%
  \selectfont}%
\fi\endgroup%
{\renewcommand{\dashlinestretch}{30}
\begin{picture}(4608,4623)(0,-10)
\thicklines
\put(2304.000,2219.625){\arc{3468.750}{4.1154}{5.3094}}
\put(2641.500,991.500){\arc{1277.204}{3.0828}{4.6536}}
\put(1966.500,991.500){\arc{1277.204}{4.7711}{6.3419}}
\put(2304.000,2465.413){\arc{1777.174}{1.9152}{7.5096}}
\put(2304.000,2484.000){\arc{3060.000}{5.7932}{9.9147}}
\path(2004,54)(2004,954)
\path(2604,54)(2604,954)
\path(3279,3654)(4179,4554)
\path(3654,3204)(4554,4104)
\path(1329,3654)(429,4554)
\path(954,3204)(54,4104)
\end{picture}
}}\hspace{-6pt}}
\newcommand{\VOneLoopXOX} {\hspace{-6pt}\raisebox{-24pt}{\setlength{\unitlength}{0.00016667in}
\begingroup\makeatletter\ifx\SetFigFont\undefined%
\gdef\SetFigFont#1#2#3#4#5{%
  \reset@font\fontsize{#1}{#2pt}%
  \fontfamily{#3}\fontseries{#4}\fontshape{#5}%
  \selectfont}%
\fi\endgroup%
{\renewcommand{\dashlinestretch}{30}
\begin{picture}(4608,4623)(0,-10)
\thicklines
\put(3822.750,3100.875){\arc{1551.272}{2.6779}{3.9355}}
\put(937.929,3126.321){\arc{1313.592}{5.3502}{6.8859}}
\put(2304.000,2360.250){\arc{1987.500}{4.1558}{5.2690}}
\put(2304.000,2439.000){\arc{3030.000}{1.7701}{7.6546}}
\put(2304.000,2437.594){\arc{1767.187}{5.9170}{9.7910}}
\put(1366.500,3691.500){\arc{1277.204}{0.8685}{2.2731}}
\put(3241.500,3691.500){\arc{1277.204}{0.8685}{2.2731}}
\path(2004,54)(2004,954)
\path(2604,54)(2604,954)
\path(3279,3654)(4179,4554)
\path(3654,3204)(4554,4104)
\path(1329,3654)(429,4554)
\path(954,3204)(54,4104)
\end{picture}
}}\hspace{-6pt}}
\newcommand{\VOneLoopXXX} {\hspace{-6pt}\raisebox{-24pt}{\setlength{\unitlength}{0.00016667in}
\begingroup\makeatletter\ifx\SetFigFont\undefined%
\gdef\SetFigFont#1#2#3#4#5{%
  \reset@font\fontsize{#1}{#2pt}%
  \fontfamily{#3}\fontseries{#4}\fontshape{#5}%
  \selectfont}%
\fi\endgroup%
{\renewcommand{\dashlinestretch}{30}
\begin{picture}(4608,4623)(0,-10)
\thicklines
\put(3822.750,3100.875){\arc{1551.272}{2.6779}{3.9355}}
\put(937.929,3126.321){\arc{1313.592}{5.3502}{6.8859}}
\put(2304.000,2360.250){\arc{1987.500}{4.1558}{5.2690}}
\put(1366.500,3691.500){\arc{1277.204}{0.8685}{2.2731}}
\put(3241.500,3691.500){\arc{1277.204}{0.8685}{2.2731}}
\put(2641.500,991.500){\arc{1277.204}{3.0828}{4.6536}}
\put(1966.500,991.500){\arc{1277.204}{4.7711}{6.3419}}
\put(2395.570,2411.267){\arc{1619.117}{5.8460}{7.5936}}
\put(2212.430,2411.267){\arc{1619.118}{1.8312}{3.5787}}
\put(2304,2454){\ellipse{3000}{3000}}
\path(2004,54)(2004,954)
\path(2604,54)(2604,954)
\path(3279,3654)(4179,4554)
\path(3654,3204)(4554,4104)
\path(1329,3654)(429,4554)
\path(954,3204)(54,4104)
\end{picture}
}}\hspace{-6pt}}

\newcommand{\VacuumOneLoop} {\hspace{-6pt}\raisebox{-20pt}{\setlength{\unitlength}{0.00016667in}
\begingroup\makeatletter\ifx\SetFigFont\undefined%
\gdef\SetFigFont#1#2#3#4#5{%
  \reset@font\fontsize{#1}{#2pt}%
  \fontfamily{#3}\fontseries{#4}\fontshape{#5}%
  \selectfont}%
\fi\endgroup%
{\renewcommand{\dashlinestretch}{30}
\begin{picture}(3676,3689)(0,-10)
\thicklines
\put(1838,1837){\ellipse{3600}{3600}}
\put(1838,1837){\ellipse{2400}{2400}}
\end{picture}
}
}\hspace{-6pt}}
\newcommand{\VacuumOneLoopTrace} {\hspace{-6pt}\raisebox{-20pt}{\setlength{\unitlength}{0.00016667in}
\begingroup\makeatletter\ifx\SetFigFont\undefined%
\gdef\SetFigFont#1#2#3#4#5{%
  \reset@font\fontsize{#1}{#2pt}%
  \fontfamily{#3}\fontseries{#4}\fontshape{#5}%
  \selectfont}%
\fi\endgroup%
{\renewcommand{\dashlinestretch}{30}
\begin{picture}(3703,3708)(0,-10)
\thicklines
\put(1852.000,1851.500){\arc{3625.000}{4.8787}{10.8293}}
\put(1852.000,1857.750){\arc{2437.500}{4.9611}{10.7469}}
\path(1552,3639)(1552,3039)
\path(2152,3639)(2152,3039)
\end{picture}
}
}\hspace{-6pt}}
\newcommand{\VacuumOneLoopPhoton} {\epsFig{-20pt}{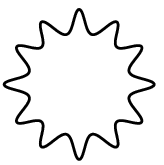}}
\newcommand{\VacuumFiveLoopPhoton} {\epsFig{-38pt}{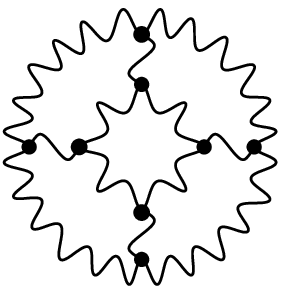}}

\newcommand{\PTwoLoopNonPlanar} {\epsFig{-26pt}{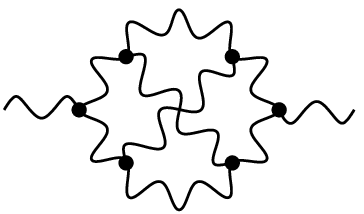}}
\newcommand{\TwoPointOneLoopOX} {\hspace{-6pt}\raisebox{-14pt}{\setlength{\unitlength}{0.00016667in}
\begingroup\makeatletter\ifx\SetFigFont\undefined%
\gdef\SetFigFont#1#2#3#4#5{%
  \reset@font\fontsize{#1}{#2pt}%
  \fontfamily{#3}\fontseries{#4}\fontshape{#5}%
  \selectfont}%
\fi\endgroup%
{\renewcommand{\dashlinestretch}{30}
\begin{picture}(5208,3048)(0,-10)
\thicklines
\path(204,1816)(204,1816)(204,1216)
\path(5004,1816)(5004,1216)
\path(54,1666)(354,1366)
\path(354,1666)(54,1366)
\path(4854,1666)(5154,1366)
\path(5154,1666)(4854,1366)
\put(2592.587,1516.000){\arc{1777.174}{0.3444}{5.9388}}
\put(2626.115,1516.000){\arc{2955.768}{3.3460}{9.2204}}
\put(4066.500,1178.500){\arc{1277.204}{3.2003}{4.6536}}
\put(4066.500,1853.500){\arc{1277.204}{1.6296}{3.0828}}
\path(1179,1816)(204,1816)
\path(1179,1216)(204,1216)
\path(5004,1816)(4029,1816)
\path(5004,1216)(4029,1216)
\end{picture}
}}\hspace{-6pt}}

\begin{document}

\thispagestyle{empty}

\begin{flushright}
TAUP-2624-2000\\
{\tt hep-th/0003235}
\end{flushright}

\title{The $U(1)$s in the Finite $N$ Limit of Orbifold Field Theories}
\bigskip
\begin{center}
{\LARGE\bf
The $U(1)$s in the Finite $N$ Limit of Orbifold Field Theories
}
\end{center}

\author{Ehud Fuchs}
\bigskip
\begin{center}
{\bf Ehud Fuchs}\footnote{\tt udif@tau.ac.il}
\end{center}

\begin{center}
{\em School of Physics and Astronomy\\
  Beverly and Raymond Sackler Faculty of Exact Sciences\\
  Tel Aviv University, Ramat Aviv, 69978, Israel
}
\end{center}

\bigskip
\begin{abstract}
We study theories generated by orbifolding the $\N=4$ super conformal
$U(N)$ Yang Mills theory with finite $N$, focusing on the r\^ole of the
remnant $U(1)$ gauge symmetries of the orbifold process.
It is well known that the one loop beta functions of the non abelian $SU(N)$
gauge couplings vanish in these theories.
It is also known that in the large $N$ limit the beta functions vanish to all
order in perturbation theory.
We show that the beta functions of the non abelian $SU(N)$ gauge couplings
vanish to two and three loop order even for finite $N$.
This is the result of taking the abelian $U(1)$ of $U(N)=SU(N)\otimes U(1)$
into account.
However, the abelian $U(1)$ gauge couplings have a non vanishing beta
function. Hence, those theories are not conformal for finite $N$.
We analyze the renormalization group flow of the orbifold theories,
discuss the suppression of the cosmological constant and tackle the
hierarchy problem in the non supersymmetric models.
\end{abstract}

\vfill
March 2000

\setcounter{page}{0}
\newpage
\setcounter{footnote}{0}

\thispagestyle{empty}
\tableofcontents
\setcounter{page}{0}
\newpage

\section{Introduction}

Supersymmetric Conformal Field Theories (SCFT) in the large $N$ limit
have been extensively studied and are very well understood.
Both the hierarchy problem and the cosmological constant problem are solved
in SCFT.
Unfortunately, we live in a non-supersymmetric non-conformal finite $N$ world.
Orbifolds of SCFT give us an opportunity to study non-SCFT using our
knowledge of SCFT and, hopefully, without loosing all the properties
of SCFT.
In this paper, we analyze orbifolds of SCFT with finite $N$, focusing
on the r\^ole of the $U(1)$ gauge symmetries that the orbifold process
leaves us with.

The large $N$ limit was first introduced by 't Hooft \cite{'tHooft:74}
who taught us that planar diagrams dominate the amplitudes of $U(N)$
gauge theories in the large $N$ limit.
He also noticed the analogy between the topologies of Feynman diagrams
and the topologies of strings of the dual string model.

More recently, Maldacena conjectured \cite{Maldacena:98} that there is
a correspondence between type IIB string theory on $AdS_5\times S^5$ and
four dimensional $\N=4$ $U(N)$ SCFT.
In the large $N$ limit it is a correspondence between IIB supergravity
and $\N=4$ SCFT. (For a review and references see \cite{Aharony:99}.)

In ``The Wall of the Cave'' \cite{Polyakov:98} Polyakov suggested that
non supersymmetric non conformal field theories should be described by
type 0 string theory.
The problem with type 0 string theory is that it has a tachyon in the
closed string sector.
Klebanov and Tseytlin showed in \cite{Klebanov:98} that the coupling of
the tachyon to the R-R fields shifts the effective mass of the tachyon
and can cure its instability.

In type 0B there is a doubling of the R-R sector.
Specifically, the five-form field strength $F_5$ is unconstrained, giving
rise to electric and magnetic D3 branes.
In \cite{Klebanov:99} the field theory living on $N$ electric and
$N$ magnetic D3 branes was first analyzed. It is an $SU(N)\otimes SU(N)$
non supersymmetric theory. The gauge coupling one loop $\beta$ function
is zero and the two loops $\beta$ function vanishes in the large
$N$ limit, suggesting that in the large $N$ limit this is a
non-supersymmetric conformal field theory.

We noticed that the two loop $\beta$ function also vanishes for finite
$N$ if a diagonal
$U(1)$ gauge field with a matching $U(1)$ scalar is included in the model.
This observation was the trigger to this paper.

The inclusion of the $U(1)$ fields makes the $SU(N)$ two loop $\beta$
function vanish. However, the $\beta$ function of the $U(1)$ gauge is non
vanishing already at one loop. Hence, the theory is not conformal
for finite $N$.

The $U(N)\otimes U(N)$ model is a $\Z_2$ orbifold of $\N=4$ $U(2N)$
super Yang Mills where $\Z_2$ is in
the center of the $SU(4)_R$ symmetry \cite{Klebanov:99,Nekrasov:99}.
This leads to the possibility that including the $U(1)$ fields in general
$\N=4$ orbifolds makes the two loop $\beta$ function vanish.

Orbifolds in the AdS/CFT correspondence where first considered in
\cite{Kachru:98}. In string theory the orbifold acts on the
$SO(6)\sim SU(4)$ isometry group of $S^5$. In field theory the 
orbifold acts on the $SU(4)_R$ symmetry.

Na\"{\i}vely, one expects that the orbifolds will have no effect on the other
symmetries of the theory.
On the string theory side, this means that the isometry group of $AdS_5$,
i.e. $SO(4,2)\sim SU(2,2)$, remains intact.
On the field theory side, this means that the conformal group
$SO(4,2)$ is not broken, leading to a conformal field theory.

However, the na\"{\i}ve expectation is not realized.
The one loop $\beta$ function of the gauge coupling does
vanish \cite{Lawrence:98}, But the higher loop corrections vanish only in
the large $N$ limit \cite{Vafa:98,Bershadsky:98}.
The source of the large $N$ requirement in orbifolds is not obvious from
the field theory point of view since the original $\N=4$ theory is conformal
also for finite $N$.

We claim that taking
``The $U(1)$s in the finite $N$ limit of Orbifold Field Theories''
into account is required for the understanding of the orbifolded theories.
The $U(1)$s can be ignored by setting their couplings to zero,
yet the vanishing of the two loop $\beta$ functions when the $U(1)$s are
taken into account signifies their r\^ole in the orbifold theories.

Orbifold theories with finite $N$ were already analyzed in the literature
but without taking the $U(1)$ factors into account.
In \cite{Frampton:98}, the conditions for the canceling of the two loop
$\beta$ functions were considered.
In \cite{Frampton:99} it was suggested that softly broken conformal
symmetry could solve the hierarchy problem.
In \cite{Csaki:99} the RG flow of the orbifold theories was analyzed.

We start our analysis of finite $N$ theories by presenting in
section \ref{sec:Unitary} the
double line notation for finite $N$. We find a subset of diagrams that have
no subleading corrections in $N$, and entitle them as ``calculable''.

In section \ref{sec:Orbifolds} we present the double line notation for
orbifold theories.
We claim that it is natural to choose all the coupling in the orbifold
theory equal
and introduce the concept of a natural line on which all the couplings
are equal and related to the original $\N=4$ coupling.
We prove the vanishing of the $\beta$ function up to three loops for
finite $N$ on the natural line.
Our proof is based on the proof for the large $N$
limit \cite{Vafa:98,Bershadsky:98}
combined with the fact that all diagrams up to three loops are ``calculable''.

In subsection \ref{sec:Hierarchy}  we discuss the scalar mass corrections
that vanish up to three loops for most of the scalars. This helps to solve
the hierarchy problem in orbifolds that do not have $U(1)$ scalars.
In subsection \ref{sec:Cosmological} we discuss the vacuum bubble diagrams
that vanish up to four loops. This could have solved the cosmological
constant problem if it were not for the running of the $U(1)$ couplings.
However, it still leads to a suppressed cosmological constant relative to
general non supersymmetric field theories.
In subsection \ref{sec:Anomaly} we discuss the $U(1)$ anomalies
in the chiral orbifolds.

Since the orbifold theories are not conformal for finite $N$,
an analysis of the renormalization group flow is in order.
This is done in section \ref{sec:RGFlow}.
We start with the easiest case when the orbifold projection leaves us
with an $\N=2$ supersymmetry. In this case the $U(1)$ fields are decoupled
from the $SU(N)$ fields leaving the $SU(N)$ theory conformal,
but strictly speaking, those theories are not conformal because of
the running of the $U(1)$ couplings.

For the $\N=1$ orbifolds we use the arguments of \cite{Leigh:95} to analyze
the manifold of fixed point. We find only the fixed line found in
\cite{Kachru:98} when the $U(1)$ fields decouple and show that the natural
line flows to the fixed line in the IR.

For the non-supersymmetric orbifolds, the lack of any non renormalization
theorems limits our results to what we can directly calculate.
We calculate the effective scalar potential to one loop order in an attempt
to check the validity of the orbifold theory.
We also calculate the $\beta$ functions to determine the RG flow
of the model.

In section \ref{sec:Summary} we summarize our results and 
discuss the prospects of generalizing the proof of the vanishing
of the $\beta$ function to all orders in perturbation theory.
We also point out
some open issues and related topics not pursued in this paper.

\section{The Unitary Group in Double Line Notation}
\label{sec:Unitary}

\subsection{Elementary Group Theory}

We start with a short presentation of the unitary group in order to
introduce the double line notation. The double line notation
introduced by 't Hooft \cite{'tHooft:74} gives the leading order behavior
in $N$. We present a notation that gives exact results including
subleading terms in $N$. Our notation closely resembles Cvitanovi\'c's
birdtracking notation \cite{Cvitanovic:76,Cvitanovic:84}.

The unitary group $U(N)$ is the group of unitary transformations on a
vector (quark) $q$ with $N$ complex components, leaving
$\bar{q}{q}=\delta^i_j q_i q^j$ invariant. The Kronecker delta is the
projection operator (propagator) of the defining (fundamental)
representation
\begin{eqnarray}
\s{i}\PFund\s{j} &=& \delta^i_j \ .
\label{eqn:PFund}
\end{eqnarray}

All invariant tensors can be constructed by products of Kronecker deltas.
We are mainly interested in the adjoint representation, since all
matter in the $\N=4$ SYM model is in this representation.
The adjoint representation is constructed from a quark-antiquark
state. There are two invariant tensors for the quark-antiquark
state, the identity $\Ident$ and the trace $\Trace$
\begin{eqnarray}
\s{a}\PPhoton\s{b} \Rightarrow \left\{
\begin{array}{rcl}
\Ident &=& ^{j_1}_{i_1}\PIdent^{j_2}_{i_2}
    = \delta^{i_1}_{i_2}\delta^{j_2}_{j_1} = \delta^{ab} \\
\Trace &=& ^{j_1}_{i_1}\PTrace^{j_2}_{i_2}
    = \delta^{i_1}_{j_1}\delta^{i_2}_{j_2} \ .
\end{array}\right.
\end{eqnarray}
Where $a,b=1\ldots N^2$ or in other words $a=(^i_j)$.
The eigenvalues of the trace matrix can be calculated using
the trace tensor equation
\begin{eqnarray*}
\Trace^2 = \PTraceSqr = N \PTrace =  N\Trace \ .
\end{eqnarray*}
The roots of the equation are $\lambda_1=N$ and $\lambda_2=0$.
With each root we can associate a projection operator
$P_i=\frac{\Trace-\lambda_j\Ident}{\lambda_i-\lambda_j}$
\begin{eqnarray}
P_{SU(N)} = \frac{\Trace-N\Ident}{0-N} &=& \PIdent-\smallfrac{1}{N}\PTrace \ ,
\label{eqn:PTraceless}\\
P_{U(1)} = \frac{\Trace-0\Ident}{N-0} &=& \smallfrac{1}{N}\PTrace \ .
\label{eqn:PTrace}
\end{eqnarray}
Those projection operators are orthonormal, $P_iP_j=\delta_{ij}P_j$, and
complete, $\sum P_i=\Ident$,
giving us the $SU(N)$ and the $U(1)$ propagators.

The generator of the defining representation (the quark-antiquark
gluon vertex) is
\begin{eqnarray}
(T^a)^i_j &=& \VQuark = c\VFund \ .
\label{eqn:VFund}
\end{eqnarray}
where $c$ is an overall normalization set by the Dynkin index of the
fundamental representation
\begin{eqnarray}
\tr[T^aT^b] &=& C(F)\delta^{ab} = \nonumber\\
c^2\POneLoopFund &=& C(F)\PIdent \\
&\Rightarrow& c^2=C(F) \ .\nonumber
\end{eqnarray}

The adjoint generator is
\begin{eqnarray}
(G^a)^{bc}=if^{abc} =
   \raisebox{12pt}{$\s{a}$}\mathop{\VPhoton}_b\raisebox{12pt}{$\s{c}$} =
   \sqrt{C(F)}\left(\VAdjoint-\VAdjointX\right) \ ,
\label{eqn:AdjGen}
\end{eqnarray}
where we choose the convention that indices are marked counterclockwise.
The antisymmetric form of the adjoint generator is required in order to
satisfy the Lie algebra for the fundamental generators,
\begin{eqnarray}
\left[T^a,T^b\right] &=& if^{abc}T^c \ ,\nonumber\\
\tChannelFund-\uChannelFund &=& \sChannelFund-\sChannelFundX \ .
\end{eqnarray}
It is easy to check that \eqref{eqn:AdjGen} also satisfies the Lie algebra for
the adjoint generators,
\begin{eqnarray}
[G^a,G^b]&=&if^{abc}G^c \ ,\nonumber\\
\tChannelPhoton-\uChannelPhoton &=& \sChannelPhoton \ .
\label{eqn:AdjAlgebra}
\end{eqnarray}

In order to calculate Feynman diagrams with gauge invariant external sources
we also need the one loop vacuum bubble diagram
\begin{eqnarray}
\VacuumOneLoopPhoton=
\VacuumOneLoop-\smallfrac{1}{N}\VacuumOneLoopTrace=
N^2-1=d(G) \ .
\label{eqn:vacuum}
\end{eqnarray}

Now we have the tools to calculate the group
factor of any Feynman diagram. Just replace each vertex with the two
vertices in \eqref{eqn:AdjGen}, each $SU(N)$ propagator with \eqref{eqn:PTraceless}
and each $U(1)$ propagator with \eqref{eqn:PTrace}, and sum all the
diagrams. For a Feynman diagram with $V$ vertices and $P$ $SU(N)$
propagators one needs to sum $2^{V+P}$ diagrams. The number of
diagrams one needs to sum can be reduced by using the fact that the
$U(1)$ propagator decouples from the adjoint vertex \eqref{eqn:AdjGen}
\begin{eqnarray}
\PTraceVAdjoint-\PTraceVAdjointX=0 \ .
\label{eqn:U1Decoupling}
\end{eqnarray}
Consequently, we can replace each $SU(N)$ propagator with the identity
propagator $\Ident$.\footnote{
Digressing to non commutative geometry, we point out that in the double
line notation it is manifest that the $U(1)$ in non commutative geometry
does not decouple from the adjoint vertex because we need to add
different phases to each diagram\cite{Bigatti:99}
\begin{eqnarray}
e^{ip_1\wedge p_2}\PTraceVAdjoint-e^{ip_2\wedge p_1}\PTraceVAdjointX \neq 0\ .
\label{eqn:U1NonComm}
\end{eqnarray}
}

The calculation of the Dynkin index for the adjoint representation
comes from the group factor of the one loop correction to the two point
function
\begin{eqnarray}
&\POneLoopOO-\POneLoopXO-\POneLoopOX+\POneLoopXX = &\nonumber\\
&2N\left(\PIdent-\smallfrac{1}{N}\PTrace\right) &
\label{eqn:PAdjOneLoop}\\
&\Rightarrow C(G)=2NC(F) \ .&\nonumber
\end{eqnarray}
We see that although we used $U(N)$ propagators, we actually calculated
the Dynkin index for the $SU(N)$ propagator. This is a result of the
$U(1)$ decoupling \eqref{eqn:U1Decoupling}. The Dynkin index for the
$U(1)$ propagator is zero.

It is useful (and easy) to calculate the group factor of the one loop
correction to the three point function
\begin{eqnarray}
&\VOneLoopOOO-3\times\left(\VOneLoopOXO-\VOneLoopXOX\right)-\VOneLoopXXX
=& \nonumber\\
&N\left(\VAdjoint-\VAdjointX\right) \ . &
\label{eqn:VAdjOneLoop}
\end{eqnarray}
The "$3\times$" stands for the three possible permutation of each
diagram. Each permutation results in a different diagram, so "$\times$"
can not be treated as the multiplicity of the diagram, but this does not
matter since the term in the brackets is zero anyway.
Equation \eqref{eqn:VAdjOneLoop} tells us that the group factor of the
one loop correction to the vertex \eqref{eqn:AdjGen} is
$NC(F)=\frac{1}{2}C(G)$.

\subsection{'t Hooft Large $N$ Limit}

Before proceeding to higher loop diagrams, we wish to recall 't Hooft
results for the large $N$ limit \cite{'tHooft:74}.
Using the double line notation we can get the $N$ dependence of a
Feynman diagram with adjoint fields from topological considerations.
A connected diagram with
$V=V_3+V_4$ vertices, $E=\frac{1}{2}(3V_3+4V_4)$ edges (propagators)
and $F$ faces (closed lines in the double line notation) has a
group coefficient proportional to
\begin{eqnarray}
g_{YM}^{V_3+2V_4}N^{F}=\lambda^{E-V}N^{\chi} \ ,
\label{eqn:planar}
\end{eqnarray}
where $\chi\equiv V-E+F=2-2g$ is the Euler characteristic and $g$ is the
genus of the surface defined by the double line diagram with all the
faces shrunk to a point.
Each Feynman diagram is translated into a number of diagrams in
the double line notation which can have different genera.
The leading $N$ contribution comes from the double line
diagrams with the minimal genus. The 't Hooft limit is defined by taking
$N$ to infinity while leaving $\lambda=g^2N$ fixed.

\subsection{``Calculable'' Diagrams}

For some diagrams we can calculate not only the leading $N$ contribution,
but the exact $N$ dependence using \eqref{eqn:AdjAlgebra},
\eqref{eqn:PAdjOneLoop} and \eqref{eqn:VAdjOneLoop}.
We will refer to diagrams that can be thus calculated
as ``calculable'' (all diagrams are calculable but the ``calculable'' ones
are easily so).
We now show generally that 
{\em any ``calculable'' $L$-loop Feynman diagram of adjoint fields has a
  group factor proportional to
}
\begin{eqnarray}
g_{YM}^{V_3+2V_4} N^{L-1}(N^2-1)=\lambda^{E-V}(N^2-1) \ .
\label{eqn:Lloop}
\end{eqnarray}
{\em  with no subleading corrections in $N$.
}
Here $L$ is the number of loops in the Feynman diagram. It can be defined as
the number of momentum loops needed to be integrated over in the vacuum
bubble diagram.
The factor of $(N^2-1)$ comes from the one loop vacuum
bubble diagram \eqref{eqn:vacuum} that has $L=1$.
For every two
3-point vertices and for every 4-point vertex we add to the diagram
we get one more loop, therefore $L=1+\frac{1}{2}V_3+V_4$.

When calculating the group factor of the diagram, each time
\eqref{eqn:PAdjOneLoop} or \eqref{eqn:VAdjOneLoop} is used, a factor of $N$
is added and a loop is removed. \eqref{eqn:AdjAlgebra} does not change neither
the power of $N$ nor the number of loops. Hence, after using
\eqref{eqn:PAdjOneLoop} or
\eqref{eqn:VAdjOneLoop} $L-1$ times, we get the one loop vacuum bubble
diagram \eqref{eqn:vacuum} with a factor of $N^{L-1}$.

We assumed that each four point vertex has the group structure of two
3-point vertices. This is not true in general, but it is true for
the $\N=4$ model.

Equation \eqref{eqn:planar} tells us that the leading order
contribution in $N$ comes from the diagram with the minimal genus.
Equation \eqref{eqn:Lloop} tells us that for ``calculable'' diagrams the
exact $N$ dependence is $(N^2-1)$.
The two equations can match only for diagrams with genus zero.
Hence we conclude that
{\em non-planar ``calculable'' diagrams have a vanishing group factor.
}
For example the non planar two loop correction to the photon is
``calculable'' which means that
\begin{eqnarray}
\PTwoLoopNonPlanar=0 \ .
\label{eqn:NonPlanar}
\end{eqnarray}
The first ``incalculable'' vacuum diagram is the five loop vacuum
bubble diagram
\begin{eqnarray}
&\VacuumFiveLoopPhoton=&
\label{eqn:5Loops}\\
&2C(F)^4(N^4+12N^2)\VacuumOneLoopPhoton=
2C(F)^4(N^4+12N^2)d(G) \ .& \nonumber
\end{eqnarray}
It is ``incalculable'' because in each loop there are four vertices and
we can not change this with the use of \eqref{eqn:AdjAlgebra}.
We calculated the group factor of this ``incalculable'' diagram by
summing up all the $2^V=256$ double line diagrams.

The first ``incalculable'' propagator diagram is the four loop diagram
obtained by cutting out a propagator from \eqref{eqn:5Loops}.
The first ``incalculable'' three vertex diagram is the three loop
diagram obtained by cutting out a vertex from \eqref{eqn:5Loops}.

\section{Orbifolds in the Double Line Notation}
\label{sec:Orbifolds}

\subsection{The Orbifold Process}

The orbifold of $\N=4$ SYM is defined by a discrete subgroup
$\Gamma$ of the global $R$ symmetry group $SU(4)_R$.
The action of the orbifold on the gauge group $U(|\Gamma|N)$ is
defined by the $\gamma$ matrices
\begin{eqnarray*}
g\in\Gamma &:& g\rightarrow\gamma_g \ ,
\end{eqnarray*}
where $\gamma_g$ are $(|\Gamma|\times|\Gamma|)\otimes\One_{N\times N}$
matrices with $\gamma_1=\One_{|\Gamma|\times|\Gamma|}\otimes\One_{N\times N}=
\One_{|\Gamma|N\times |\Gamma|N}$.
The orbifold breaks the gauge group into $U(N)^{|\Gamma|}$.
For simplicity we assumed that all irreducible representations of $\Gamma$
are one dimensional, namely that $\Gamma$ is abelian.
For a general discrete group $\Gamma$ with irreducible representations
labeled $r_i$,
the orbifold breaks the gauge group $U(\sum_i d_i N)$
into the $\bigotimes_i U(d_i N)$ gauge group where $d_i=\dim r_i$.

The cancellation of tadpoles in string theory imposes that the
representation of $\Gamma$ has to be regular \cite{Vafa:98},
meaning $\tr[\gamma_g]=0$ $\forall g\neq 1$. The regularity of $\Gamma$
guarantees the cancellation of the $SU(N)$ gauge anomalies, though it is not
a necessary condition.

The spectrum of the orbifold theory is defined by the projection operator
\cite{Bershadsky:98}
\begin{eqnarray}
P=\frac{1}{|\Gamma|}\sum_{g\in\Gamma}
  r_g\otimes\gamma_g^{\dagger}\otimes\gamma_g \ ,
\label{eqn:OrbifoldProj}
\end{eqnarray}
where $r_g$ is the representation of the projected field under
the $SU(4)_R$ symmetry group. The projection leaves only fields that
are invariant under the orbifold $\Gamma$, i.e.
\begin{eqnarray*}
A_{\mu}=\gamma_g^{\dagger}A_{\mu}\gamma_g && \forall g\in \Gamma \ ,\\
\phi_I=\gamma_g^{\dagger}(r^6_g)^J_I\phi_J\gamma_g && \s{I,J=1\ldots6} \ ,\\
\psi_I=\gamma_g^{\dagger}(r^4_g)^J_I\psi_J\gamma_g
&& \s{I,J=1\ldots4} \ .
\end{eqnarray*}

For the double line notation we split the fundamental propagator
\eqref{eqn:PFund} into $|\Gamma|$ parts
\begin{eqnarray}
\delta^i_j &=& \delta^k_l\delta^{i_k}_{j_l} 
                   \ \ \ \ \s{k,l=0\ldots|\Gamma|-1} \ ,\nonumber\\
\s{i}\PFund\s{j} &=& \diag(\s{i_0}\PFundI\s{j_0},
\s{i_1}\PFundA\s{j_1},\s{i_2}\PFundB\s{j_2},\ldots) \ .
\end{eqnarray}
The action of the projection operator \eqref{eqn:OrbifoldProj} on the
adjoint propagator depends on
the group element $g$ under which the propagator transforms,
\begin{eqnarray}
g_0=1:\ 
    \delta^l_k(\Ident-\smallfrac{1}{|\Gamma|N}\Trace)
        &\eqneq&
    \diag(\PIdentII-\smallfrac{1}{N}\PTraceII, \nonumber\\
   &&\phantom{\diag}\ \PIdentAA-\smallfrac{1}{N}\PTraceAA,\ldots) \ ,\\
    \delta^l_k(\smallfrac{1}{|\Gamma|N}\Trace)
        &\eqneq&
    \diag(\smallfrac{1}{N}\PTraceII,\smallfrac{1}{N}\PTraceAA,\ldots) \ ,\\
g_{k'}=e^{\frac{2\pi i}{|\Gamma|}k'}:\ 
    \delta^l_{k+k'}(\Ident-\smallfrac{1}{|\Gamma|N}\Trace)
        &\eqneq&
    \offdiag_{k'}(\PIdentIA,\PIdentAB,\ldots) \ ,
\label{eqn:BiFundProp}
\end{eqnarray}
where $k'=1\ldots|\Gamma|-1$.
The $\Trace$ propagator is multiplied by a factor of $|\Gamma|$
because
$\delta^l_k\delta^k_l\delta^{i_k}_{j_l}=|\Gamma|\delta^{i_k}_{j_l}$.
We see that fields that are invariant under $SU(4)_R$ are in adjoint
representations of 
${\displaystyle\left(\vphantom{1^1}SU(N)\otimes U(1)\right)}^{|\Gamma|}$,
while non-invariant fields are in bifundamental representations.
$\offdiag_{k'}$ means that the nonzero elements are shifted $k'$ places
off the diagonal.

In \eqref{eqn:BiFundProp} we assumed that the orbifold group is
$\Gamma=\Z_{|\Gamma|}$. For other groups we would have a different
representation for $g_{k'}$ and would get the bifundamental
fields in different permutations.

The embedding of $\Z_{|\Gamma|}$ in $SU(4)_R$ can be specified using four
integer weights $(k_1,k_2,k_3,k_4)$ describing how $\Z_{|\Gamma|}$ operates
on the fundamental representation ${\bf 4}$ of $SU(4)_R$. Because we
are interested in subgroups of $SU(4)$ and not of $U(4)$, we require
$k_1+k_2+k_3+k_4=0 \mod |\Gamma|$.
Consequently, the embedding of $\Z_{|\Gamma|}$ is
parameterized by three integer numbers. The transformation of the
antisymmetric representation ${\bf 6}$ is described by the
six integers $k_{(i,j)}=k_i+k_j$, where $(i,j)$ are $(^4_2)$ unordered
pairs.

To describe a general orbifold group, not necessarily abelian, we can
use quiver diagrams \cite{Douglas:96} with a node for each irreducible
representation of $\Gamma$ ($|\Gamma|$ nodes for abelian groups).
The diagrams have
four directed links from each node describing the four fermions $(\psi)$,
four directed links going into each node $(\bar{\psi})$,
three undirected (double directed) links from each node describing the
six real scalars (three complex scalars) and one undirected link going
from each node to itself describing the gauge field in the adjoint
representation.

The effect of the orbifold on the three vertex \eqref{eqn:AdjGen} is
described by three integers that must satisfy $k_1+k_2+k_3=0 \mod |\Gamma|$
(a closed loop in the quiver language), since the vertex is
$SU(4)_R$ invariant. If we choose $k_1=-k_2\equiv k$, $k_3=0$,
we get the bifundamental generator
\begin{eqnarray}
(T^{a_l})^{i_l}_{j_l}\delta^{j_{l+k}}_{i_{l+k}} &-&
(T^{a_l})^{j_l}_{i_l}\delta^{i_{l-k}}_{j_{l-k}}= \nonumber\\
\sqrt{C(F)}\left(
\raisebox{12pt}{$^{i_{l+k}}_{i_l}$}\mathop{\VBifund}_{a_l}
\raisebox{12pt}{$^{j_{l+k}}_{j_l}$}\right. &-& \left.
\raisebox{12pt}{$^{i_l}_{i_{l-k}}$}\mathop{\VBifundX}_{a_l}
\raisebox{12pt}{$^{j_l}_{j_{l-k}}$}
\right) \ .
\label{eqn:BiGen}
\end{eqnarray}
We can read from \eqref{eqn:BiGen} that an adjoint in $SU(N)_l$ couples
to a couple of bifundamentals, $(N_l,\overline{N}_{l+k})$ and
$(N_{l-k},\overline{N}_l)$. For $k_1,k_2,k_3\neq0$ we get vertices of three
bifundamental. Those vertices exist only for some orbifolds, those that
have a triangle with all vertices on different nodes of the quiver diagram.

It is obvious that the $U(1)$ factors do not decouple any more,
\begin{eqnarray}
\PTraceVBifund-\PTraceVBifundX \neq 0 \ .
\label{eqn:U1Coupling}
\end{eqnarray}
The $|\Gamma|$ $U(1)$ factors of the orbifold theory are not independent,
as the sum over the $U(1)$ charges for each field is zero.
This is a consequence of the decoupling of the $U(1)$ from the original
$U(|\Gamma|N)$ theory \eqref{eqn:U1Decoupling}.

The Dynkin index of the bifundamental representation is
\begin{eqnarray}
&\POneLoopBifund-\smallfrac{1}{N}\POneLoopBifundTrace =
N\left(\PIdent-\smallfrac{1}{N}\PTrace\right) & \nonumber\\
&\Rightarrow C^{SU(N)}(B)=NC(F) \ ,&
\label{eqn:POneLoopBiFund}\\
&\smallfrac{1}{N}\POneLoopBifundTrace =
N\left(\smallfrac{1}{N}\PTrace\right) & \nonumber\\
&\Rightarrow C^{U(1)}(B)=NC(F) \ .&
\label{eqn:POneLoopBiFundU1}
\end{eqnarray}
The $N$ dependence of the $U(1)$ Dynkin index comes from the multiplicity
of the bifundamental representation.
Diagrams with one twisted vertex are projected out in the orbifold process
\eqref{eqn:OrbifoldProj}. Diagrams with two twisted vertices will give the
Dynkin index of the anti-bifundamental representation $C(\bar{B})=NC(F)$.
One can see that we no longer have the luxury of ignoring the trace
propagator $\Trace$.

The second Casimir $(T^aT^a)^i_j=C_2(F)\delta^i_j$ of the bifundamental
representation is calculated from the one loop correction to the
bifundamental propagator,
\begin{eqnarray}
&\PBifundOneLoop-\smallfrac{1}{N}\PBifundOneLoopTrace
 = \smallfrac{N^2-1}{N}\PIdentIA & \nonumber\\
&\Rightarrow C^{SU(N)}_2(B)=\smallfrac{N^2-1}{N}C(F) \ ,&
\label{eqn:PBiOneLoop}\\
&\smallfrac{1}{N}\PBifundOneLoopTrace
 = \smallfrac{1}{N}\PIdentIA & \nonumber\\
&\Rightarrow C^{U(1)}_2(B)=\smallfrac{1}{N}C(F) \ .&
\label{eqn:PBiOneLoopU1}
\end{eqnarray}

The gauge group is 
$\displaystyle\left(SU(N)\otimes U(1)\right)^{|\Gamma|}$ meaning that for
every $SU(N)$ second Casimir there is a $U(1)$ second Casimir. The
contribution of the one loop bifundamental propagator to the Feynman
diagram will alway be of the form
\begin{eqnarray}
\left(g_N^2 \smallfrac{N^2-1}{N} + g_1^2 \smallfrac{1}{N}\right)C(F) =
\left(g_N^2 N + \frac{g_1^2-g_N^2}{N} \right)C(F) \ ,
\label{eqn:C2}
\end{eqnarray}
where $g_N$ is the $SU(N)$ gauge coupling and $g_1$ is the $U(1)$
gauge coupling.
From \eqref{eqn:C2} we see that the $U(1)$ factor can be neglected in the
large $N$ limit, in the 't Hooft limit it is suppressed by a factor of
$\frac{1}{N^2}$.
We also see that if we choose $g_N=g_1$, the subleading corrections in
$N$ are canceled.
This is the natural choice since we originally had a $U(N)$
symmetry that was split by the RG flow to $SU(N)\otimes U(1)$.

\subsection{The Natural Line}

The orbifold theory has $2|\Gamma|$ gauge couplings.
In the space of gauge couplings we choose the two dimensional
manifold parametrized by $(g_N,g_1)$ for which all
the $SU(N)$ couplings are the same and all the $U(1)$ couplings are
the same. It is the natural choice since we originally had a
$U(|\Gamma|N)$ symmetry.
For the $\Z_{|\Gamma|}$ orbifold this manifold has a
$\Z_{|\Gamma|}$ symmetry and all the RG equations have a
$\Z_{|\Gamma|}$ symmetry. Accordingly, if we start in this manifold, we
will stay in it.
We also choose the Yukawa couplings to be equal to the gauge coupling
and the quartic couplings to be equal to the gauge coupling squared.
The RG flow can make those couplings different.

In the space of couplings, we choose to start the RG flow from a point
on a one dimensional manifold (line) parametrized by
the coupling $g$ to which all the couplings are equal at some renormalization
scale $\mu_N$. $g$ can be related to the coupling of the original $\N=4$
theory.
This is the natural submanifold to choose because of the $\N=4$ origin
of the orbifold theory and we will refer to this submanifold as the
{\em natural line}.

In view of the AdS/CFT correspondence, $g$ can be related to the string
coupling $g^2\sim g_s$.
There are two scales in the field theory orbifold, the regularization scale
$\Lambda$ and the renormalization scale $\mu_N$. In the AdS/CFT correspondence
the regularization scale is related to the string scale
$\Lambda\sim \frac{1}{\sqrt{\alpha'}}$, and the renormalization scale is
related to the $AdS_5$ fifth coordinate $\mu\sim U=\frac{r}{\alpha'}$.
Our model is not conformal, therefore we do not expect an AdS geometry.
The renormalization scale $\mu_N$ where all the couplings are equal, should
be related to some unique $U_N$ in the new geometry. From dimensional
consideration it should also be somehow related to the string scale.
The regularization scale in field theory is not physical, but we have to
take it into consideration when we discuss the hierarchy problem and the
cosmological constant problem.

In (\ref{eqn:POneLoopBiFund}-\ref{eqn:PBiOneLoopU1}) we calculated diagrams
obtained by orbifold projections of the one loop diagram
\eqref{eqn:PAdjOneLoop}.
We found out that their group factor is proportional to $g^2N$ with no
subleading corrections on the natural line.
The only other non trivial orbifold projection of \eqref{eqn:PAdjOneLoop} is
\begin{eqnarray}
\PBifundOneLoopYukawa=N\PIdentIA \ .
\end{eqnarray}
The vertices in this diagram can only be Yukawa vertices and
since the Yukawa couplings are equal to the gauge couplings on the natural
line, this diagram also has a group factor of $g^2N$.

The orbifold projections of the one loop
correction to the three point vertex \eqref{eqn:VAdjOneLoop}
also have a group factor proportional to $g^2N$ with no subleading corrections
in $N$, when all the couplings are equal to $g$.
``Calculable'' diagrams were defined as diagrams that can be calculated
using \eqref{eqn:PAdjOneLoop}, \eqref{eqn:VAdjOneLoop} together with
\eqref{eqn:AdjAlgebra}. Hence, we conclude that
{\em ``calculable'' diagrams of orbifold theories have no subleading
  corrections in $N$ on the natural line.
}
The group factor of those diagrams is the same as in \eqref{eqn:Lloop}.

\subsection{Vanishing of the $\beta$ Functions}

In \cite{Vafa:98,Bershadsky:98} it was shown that in the large $N$
limit the correlation
functions of the orbifold theories coincide with those of $\N=4$.
This leads to the vanishing of the $\beta$ function of the
orbifold theories to all orders in perturbation theory in the large
$N$ limit.
We want to generalize the proof for finite $N$, but not to all orders,
only to orders for which all diagrams in that order are ``calculable''.

The proof in \cite{Vafa:98,Bershadsky:98} was for planar diagrams with
all external legs attached to the same boundary.
The fact that ``calculable'' diagrams have no subleading corrections
on the natural line leads to the conclusion that non planar ``calculable''
diagrams have a vanishing contribution.
For example, all orbifold projections of \eqref{eqn:NonPlanar} will vanish
on the natural line.
If the external legs are attached to different boundaries there are
several possibilities \cite{Vafa:98}
\begin{itemize}
\item For two point functions, each leg is attached to a boundary of
  itself. The color indices of the leg are traced, meaning that they
  are $U(1)$ legs. This is the source of the running of the $U(1)$
  coupling constant
\begin{eqnarray*}
\TwoPointOneLoopOX \ .
\end{eqnarray*}
\item For three point functions, one of the external
  legs must be attached to a boundary of itself. This is the source of the
  running of the Yukawa coupling of the $U(1)$ fields.
\item For four point functions, the previous argument does not
  apply and we have $SU(N)$ diagrams.
  Those diagrams have a different group structure from that of the original
  $\N=4$ quartic couplings. In orbifolds with at least $\N=1$ supersymmetry,
  the perturbative non renormalization of the superpotential guarantees that
  new quartic couplings will not be generated. In $\N=0$ orbifolds, new
  quartic couplings are generated as we shall see in the next section.
\item Five or more point functions can not be generated because we are dealing
  with renormalizable field theories.
\end{itemize}

Two point functions are ``calculable'' up to three loops, and three
point functions are ``calculable'' up to two loops, hence there are no
subleading corrections in $N$ on the natural line and the corresponding
diagrams coincide with the $\N=4$ ones. We conclude that
{\em on the natural line the two point functions of the non abelian fields
  have a zero $\beta$ function up to three loops
  and the three point functions of the non abelian fields have a zero
  $\beta$ function up to two loops.
}

The orbifold theories on the natural line are not finite because the
$\beta$ functions of the $U(1)$ couplings are non-zero already at one loop,
which may cause the orbifold theories to flow away from the natural line.
To see the flow of the $SU(N)$ couplings one has to calculate the second
derivative of the coupling, since the first derivative (the $\beta$
function) is zero on the natural line. Since the first derivative is
a function of the $U(1)$ coupling, the second derivative will depend
on the $U(1)$ $\beta$ function.

\subsection{The Hierarchy Problem}
\label{sec:Hierarchy}

Generally in field theory, the scalar
two point functions diverge quadratically leading to scalar masses of the
order of $\Lambda^2$, where $\Lambda$ is some cutoff scale. To keep the 
scalars light, mass counterterms must be very fine tuned. This is called the 
hierarchy problem.

In supersymmetric theories, the mass of the scalars is protected by
the non renormalization of the superpotential, solving the hierarchy
problem.
In non supersymmetric orbifolds, a mass counterterm is not needed for most of
the scalar fields, at least up to three loops, because of the vanishing of
the scalar two point function.
This is again a result of the matching between the $\N=4$ diagrams and
orbifold diagrams for the scalar two point functions.
This could have helped to solve the hierarchy problem for non supersymmetric
theories, but this matching does not work for the $U(1)$ scalars which are
diverging already at one loop.

The problem of the diverging $U(1)$ scalar mass is not general to all $\N=0$
orbifolds. There are $\N=0$ orbifolds with no scalars in the adjoint
representation and hence no $U(1)$ scalars, e.g., the $\Z_4$ orbifold
with weights $(1,1,1,1)$.
But in those theories the $U(1)$ gauge symmetries have anomalies as is
discussed is subsection \ref{sec:Anomaly}

The divergence of the $U(1)$ scalar mass and the cancellation of the other
scalar masses are demonstrated in subsection \ref{sec:N0} for the $Z_2$
non supersymmetric orbifold, using the effective potential formalism.

The mass of the $U(1)$ scalar depends on the regularization scheme.
If we claim that field theory is related to string theory by the AdS/CFT
correspondence, the scheme we should choose is adding massive
fields corresponding to the massive open strings between D3 branes in the
orbifolded Type IIB string theory.
Those massive fields will act as a cutoff.
We can hope that in this scheme the $U(1)$ scalars will be massless.

In \cite{Frampton:99} it was already suggested that non supersymmetric
conformal field theories with a softly broken conformal symmetry may
solve the hierarchy problem. We suggest that the $U(1)$ couplings behave
as naturally occurring soft symmetry breaking terms of the conformal
symmetry in the sense that the flow of the $U(1)$ couplings induces the
flow of the other couplings. The ``soft breaking'' parameter is
$\frac{1}{N}$.

In \cite{Csaki:99} the $U(1)$ factors were not taken into account resulting in
a mass term for all scalar fields suppressed by a factor of $\frac{1}{N}$.
There it was suggested to solve the hierarchy problem by choosing a
{\em very large} but finite $N$.
Notice that when the $U(1)$ factors are taken into account the scalar mass
vanish, but only at the renormalization scale $\mu_N$. At other scales
we will get $\Lambda^2$ contributions again.

\subsection{The Cosmological Constant}
\label{sec:Cosmological}

The vacuum energy in field theories is generally of the order of
$\Lambda^4$, where $\Lambda$ is some cutoff scale. In field theory
the constant shift of the vacuum energy is unobservable. In general
relativity the vacuum energy plays the r\^ole of the cosmological
constant. The cosmological constant is expected to be very small while
$\Lambda^4$ is very large. This is called the cosmological constant problem.

The contributions to the cosmological constant come from the zero point
functions (vacuum bubble diagrams). To one loop order, bosons and fermions of
the same mass have equal and opposite contributions to the cosmological
constant.
In orbifold theories the cosmological constant vanishes to one loop because
the number of bosons and fermions is the same.
In non supersymmetric orbifolds this would not be true without taking the
$U(1)$ factors into account.

We can go further than that in the loop expansion.
We can use the vanishing of the vacuum energy in the $\N=4$ theory to
conclude that the vacuum bubble diagrams of orbifold theories vanish on the
natural line up to four loops. (In five loops we have the ``incalculable''
bubble diagram \eqref{eqn:5Loops}.) This leads to a suppression of the
cosmological constant by a factor of $g^8$.

In \cite{Kumar:98} it was suggested that
the cosmological constant in orbifold theories vanishes in cases where one
has a non supersymmetric fixed line. In our finite $N$ non supersymmetric
models we have found no fixed line, therefore we do not expect that the
cosmological constant would vanish.

\subsection{$U(1)$ Anomalies}
\label{sec:Anomaly}

The regularity of the orbifold guarantees the cancellation of the $SU(N)$
gauge anomalies, but not of the $U(1)$ anomalies. The $U(1)$ gauge symmetries
are anomalous in the chiral orbifolds.

$Z_{|\Gamma|}$ orbifolds are not chiral if their weights are of the form
\begin{eqnarray}
(k_1,-k_1,k_2,-k_2) \ .
\label{eqn:NonChiral}
\end{eqnarray}
This means that $\N=2$ orbifolds are not chiral,
$\N=1$ orbifolds are chiral and $\N=0$ orbifolds can be either chiral or 
not chiral.

In the chiral orbifolds all the $U(1)$s are anomalous and there are no
non anomalous combination except for the trivial $U(1)$ which is completely
decoupled from the rest of the fields. An example of an anomalous orbifold
is given in subsection \ref{sec:N1}.

It was shown in \cite{Ibanez:98} that those $U(1)$ anomalies cancel by
a generalized Green-Schwarz mechanism. Fayet-Illiopoulos terms are generated
for the anomalous $U(1)$ gauge fields giving them mass of the order of the
string scale.

From the form of the non chiral orbifold \eqref{eqn:NonChiral} it can be
seen that in non chiral orbifold there are scalar $U(1)$s. This means that
in the orbifold theories with no anomalies there will be scalars that can
acquire mass as was discussed in subsection \ref{sec:Hierarchy}.

\section{The Renormalization Group Flow of Orbifold Theories}
\label{sec:RGFlow}

\subsection{General $\beta$ Functions}

Now we want to analyze the RG flow from the natural line on which the
couplings reside
at the renormalization scale $\mu_N$. We have the freedom to choose the
coupling we start with, so we can rely on perturbation theory by choosing
small $g$. We are also taking all the masses to zero at the renormalization
scale $\mu_N$.

For our analysis we calculate the gauge coupling $\beta$ function
up to two loop order because the first loop correction vanishes.
The Yukawa coupling $\beta$ function is calculated up to one loop.
We do not need to calculate the quartic coupling $\beta$ function because
the quartic coupling does not participate in the evolution of the gauge
and Yukawa couplings at the orders we are looking at.
The quartic coupling $\beta$ function will be calculated for the $\N=0$
orbifold using the effective action.

The gauge coupling beta function for a product gauge group, up to two loops,
depends on the gauge couplings $\{g\}$ and the Yukawa coupling matrices $Y$
\cite{Machacek:83},
\begin{eqnarray}
\beta_{g_k}(\{g\},Y) &\eqneq& \frac{d g_k}{d \log \mu}
            = \beta_{g_k}^{(1)}(g_k)+\beta_{g_k}^{(2)}(\{g\},Y) \ \ ,\\
\beta_{g_k}^{(1)}(g_k) &\eqneq& -\frac{g_k^3}{(4\pi)^2}\left [
\frac{11}{3}C_2(G_k)
-\frac{2}{3}\sum_{\tinybox{fermions}}C(F_k)-\frac{1}{6}\sum_{\tinybox{scalars}}{C(S_k)}
                    \right] \ ,\ \ \ \ \ \\
\beta_{g_k}^{(2)}(\{g\},Y) &\eqneq& -\frac{g_k^3}{(4\pi)^2}\left [
\frac{34}{3}\frac{g_k^2}{(4\pi)^2}C_2(G_k)^2 \right. \nonumber\\
&&-\sum_{\tinybox{fermions}}{\left(
            \sum_{l\in^{\tinybox{gauge}}_{\tinybox{groups}}}{2\frac{g_l^2}{(4\pi)^2}C_2(F_l)}
           + \frac{10}{3}\frac{g_k^2}{(4\pi)^2}C_2(G_k) \right)C(F_k)}
       \nonumber\\
&&-\sum_{\tinybox{scalars}}{\left( \sum_{l\in^{\tinybox{gauge}}_{\tinybox{groups}}}{2\frac{g_l^2}{(4\pi)^2}C_2(S_l)}
           + \frac{1}{3}\frac{g_k^2}{(4\pi)^2}C_2(G_k) \right)C(S_k)}
       \nonumber\\
&&\left.+\frac{1}{(4\pi)^2}Y_4(F)
                    \right] \ ,
\end{eqnarray}
where $G_k$ is the adjoint representation of the $k$th gauge field,
$F_k$ is the representation of the fermions under the $k$th gauge group,
$S_k$ is the representation of the scalars under the $k$th gauge group,
$(Y^S)^{F^1}_{F^2}$ is the Yukawa coupling matrices representing the
coupling between a scalar $S$ and two fermions $F^1,F^2$,
and $Y_4(F)$ is the Yukawa coupling contribution defined as
\begin{eqnarray}
Y_4(F)\delta^{ab}&=&\mathop{\tr}_{\tinybox{fermions}}\sum_{\tinybox{scalars}}
                       Y^SY^{\dagger S}T^aT^b \ .
\end{eqnarray}
The summations are over Weyl fermions and real scalars.
The Dynkin index of the fundamental representation is normalized to
$C(F)=\frac{1}{2}$.

The one loop $\beta$ function for the Yukawa coupling is \cite{Machacek:84}
\begin{eqnarray}
\lefteqn{\beta_{Y^S}=\frac{d Y^S}{d \log\mu}=} \nonumber\\
&&\frac{1}{(4\pi)^2}
\left[\vphantom{\sum_{k\in^{\tinybox{gauge}}_{\tinybox{groups}}}}
\frac{1}{2}\left(Y^{\dagger S'}Y^{S'}Y^S+Y^S Y^{S'}Y^{\dagger S'}\right)
+ 2Y^{S'}Y^{\dagger S}Y^{S'}
+ Y^{S'}\tr \left(Y^{\dagger S'}Y^S\right) \right.\nonumber\\
&&\phantom{\frac{1}{(4\pi)^2}}\left.
{} - 3\sum_{k\in^{\tinybox{gauge}}_{\tinybox{groups}}}
  \left(\vphantom{Y^{S'}}g_k^2 C_2(F^1_k)Y^S+Y^S C_2(F^2_k)\right) \right] \ .
\end{eqnarray}
The first two terms are the scalar loop corrections to the two fermion legs.
The third term is the one point irreducible scalar correction.
The fourth term is the fermion loop correction to the scalar leg.
The last two terms are the gauge bosons loop corrections to the two fermion
legs.

In the following subsections we analyze orbifold theories with
$\N=2,1$ and $0$ supersymmetries.

\subsection{$\N=2$ Orbifolds}

The orbifold leaves an $\N=2$ supersymmetry if $\Gamma\subset
SU(2)\subset SU(4)_R$. The simplest case is the $\Z_2$ orbifold with
weights $(1,1,0,0)$ that leaves the following matter content,
\begin{eqnarray*}
\begin{array}{l|c@{}c@{}c@{}c|c}
 & \multicolumn{1}{c@{\ \otimes}}{SU(N)}
 & \multicolumn{1}{c@{\ \otimes}}{SU(N)}
 & \multicolumn{1}{c@{\ \otimes}}{U(1)}
 & \multicolumn{1}{c|}{U(1)}
 & SU(2) \\
\hline
V_{N_1} & G & 1 & 0 & 0 & 1 \\
V_{N_2} & 1 & G & 0 & 0 & 1 \\
V_{1_1} & 1 & 1 & 0 & 0 & 1 \\
V_{1_2} & 1 & 1 & 0 & 0 & 1 \\
\hline\vphantom{\overline{N}^1}
H^1_2 & N & \overline{N} & \phantom{-}1 & -1 & 2
\end{array}
\end{eqnarray*}
where $V$ and $H$ are the vector and hyper multiplet of $\N=2$.
The $SU(2)$ is a global symmetry that is a remnant of the original $SU(4)_R$
symmetry. There is also the usual $SU(2)_R\otimes U(1)_R$ symmetry.

The $U(1)$ charges are written up to a normalization factor.
When calculating the group factor of Feynman diagrams, we use
the double line notation as in \eqref{eqn:POneLoopBiFundU1}
and \eqref{eqn:PBiOneLoopU1}.

The non-renormalization theorem of $\N=2$ guarantees that there can only
be one loop corrections to the perturbative $\beta$ function.
The different gauge bosons can only interact through the bifundamental
hypermultiplet and those interactions only occur at two loop order.
Consequently, the different gauge bosons do not interact.

We choose
all the $SU(N)$ couplings to be the same and all the $U(1)$ couplings to
be same, and then there are only two independent couplings, with the $\beta$
functions
\begin{eqnarray}
\beta_{g_N} &=& 0 \ ,\nonumber\\
\beta_{g_1} &=& 2\frac{g_1^3N}{(4\pi)^2} \ .
\end{eqnarray}
If we start on the natural line, we have $g_N=g_1$ and we get a theory that
is not finite because of the running of the $U(1)$ coupling constant.
However, we can choose $g_1=0$ and get a finite theory.
In other words, since the $U(1)$ gauge couplings are IR free,
we can say that the $U(1)$ decouples in the IR, and the theory is IR finite.

For a general $Z_{|\Gamma|}$ orbifold we have a $|\Gamma|$ dimensional
manifold of fixed points parameterized by the $|\Gamma|$ $SU(N)$ gauge
couplings.

\subsection{$\N=1$ Orbifolds}
\label{sec:N1}

The orbifold leaves an $\N=1$ supersymmetry if $\Gamma\subset
SU(3)\subset SU(4)_R$. The simplest case is the $\Z_3$ orbifold with weights
$(1,1,1,0)$ \cite{Kachru:98} having the following matter content,
\begin{eqnarray*}
\begin{array}{l|c@{}c@{}c@{}c@{}c@{}c|c}
 & \multicolumn{1}{c@{\ \otimes}}{SU(N)_1\!}
 & \multicolumn{1}{c@{\ \otimes}}{\!SU(N)_2\!}
 & \multicolumn{1}{c@{\ \otimes}}{\!SU(N)_3\!}
 & \multicolumn{1}{c@{\ \otimes}}{U(1)_1\!}
 & \multicolumn{1}{c@{\ \otimes}}{\!U(1)_2\!}
 & \multicolumn{1}{c|}{\!U(1)_3\!}
 & SU(3) \\
\hline
V_{N_1} & G & 1 & 1 & 0 & 0 & 0 & 1 \\
V_{N_2} & 1 & G & 1 & 0 & 0 & 0 & 1 \\
V_{N_3} & 1 & 1 & G & 0 & 0 & 0 & 1 \\
V_{1_1} & 1 & 1 & 1 & 0 & 0 & 0 & 1 \\
V_{1_2} & 1 & 1 & 1 & 0 & 0 & 0 & 1 \\
V_{1_3} & 1 & 1 & 1 & 0 & 0 & 0 & 1 \\
\hline\vphantom{\overline{N}^1}
\Phi^1_2 & N & \overline{N} & 1 & \phantom{-}1 & -1 & \phantom{-}0 & 3 \\
\Phi^2_3 & 1 & N & \overline{N} & \phantom{-}0 & \phantom{-}1 & -1 & 3 \\
\Phi^3_1 & \overline{N} & 1 & N & -1 & \phantom{-}0 & \phantom{-}1 & 3
\end{array}
\end{eqnarray*}
where $V$ and $\Phi$ are the vector and chiral multiplets of $\N=1$.
The $SU(3)$ is a global symmetry remnant of the original $SU(4)_R$
symmetry. There is also the usual $U(1)_R$ symmetry.
The superpotential is\footnote{
  The entire Lagrangian is normalized by a factor of $\frac{1}{C(F)}$.
  This is due to the use of the notation $\Phi\equiv T^a\Phi^a$ that
  contributes a factor of $C(F)$. For example,
  $\tr(D_{\mu}\Phi D^{\mu}\Phi)=\tr(T^aT^b)D_{\mu}\Phi^a D^{\mu}\Phi^b=
  C(F)D_{\mu}\Phi^a D^{\mu}\Phi^a$.
}
\begin{eqnarray}
\sqrt{2}h\sum_{k=1}^3
\tr\left([\Phi^k_{k+1},\Phi^{k+1}_{k+2}]\Phi^{k+2}_k\right) \ .
\end{eqnarray}
The trace here stands for taking the singlet representation under all gauge
groups and the $SU(3)$ global symmetry group.

Generally we can have different $h_k$ for the three summands.
However, for the sake of simplicity and naturalness, we choose
$h_k=h$, $g_N^k=g_N$ and $g_1^k=g_1$.
The RG flow will not alter this choice.
Before the orbifolding, $h=g$ would have yielded an $\N=4$ theory.

The Yukawa terms in the Lagrangian are (in components)
\begin{eqnarray}
&&\sqrt{2}g_N\sum_{k=1}^3\left(\phi^k_{k+1}\bar{\psi}^{k+1}_k\lambda^{N_k} +
  \phi^{k-1}_k\bar{\psi}^k_{k-1}\lambda^{N_k}\right) + \\
&&\sqrt{2}g_1\sum_{k=1}^3\left(\phi^k_{k+1}\bar{\psi}^{k+1}_k\lambda^{1_k} +
  \phi^{k-1}_k\bar{\psi}^k_{k-1}\lambda^{1_k}\right) +
\frac{h}{\sqrt{2}}\sum_{k=1}^3\phi^k_{k+1}\psi^{k+1}_{k+2}\psi^{k+2}_k +
\text{h.c.} \nonumber
\end{eqnarray}
where the trace over the gauge and global indices is implicit.
The coupling of the first two terms has to be equal to the gauge
coupling as a consequence of supersymmetry. The third term comes from the
superpotential.

There are non vanishing triangle anomalies for the $U(1)$ gauge symmetries
in this theory. For example
\begin{eqnarray}
U(1)_1^2U(1)_2: && \sum Q_1^2Q_2=-3N^2 \ .
\end{eqnarray}
To study the cancellation of those anomalies we need to study the effective
theory of the orbifolded string theory.
It not sufficient to study the orbifolded $\N=4$ field theory.
It was shown in \cite{Ibanez:98} that there is a generalized Green-Schwarz
mechanism that cancels those anomalies.
Fayet-Illiopoulos terms are generated to the anomalous $U(1)$s. Those FI
terms give a mass to the $U(1)$ gauge fields of the order of the string scale,
which means that in the limit $l_s\rightarrow 0$ the $U(1)$s are decoupled
from the effective field theory.

Next, we will calculate the RG flow of this theory with the $U(1)$ gauge
fields despite it being inconsistent because of the $U(1)$ anomalies. 
The purpose of the calculation is to show that the theory reaches a fix point
which is stable under fluctuations of the form of the $U(1)$ fields.

We can use the Leigh-Strassler arguments \cite{Leigh:95} to check whether 
there is a manifold of fixed points.
From $\N=1$ SUSY arguments we know \cite{Shifman:86} that the $\beta$
functions have the form
(when absorbing a factor of $4\pi$ into the couplings),
\begin{eqnarray}
\beta_h &\eqneq&
    h\left(-3+\sum\left(d(\Phi)+\smallfrac{1}{2}\gamma\right)\right)
  = \smallfrac{3}{2}h\gamma \ ,\nonumber\\
\beta_{g_N} &\eqneq& -\frac{g_N^3}{1-g_N^2C_2(G)}
    \left(3C_2(G)-\sum C(R)(1-\gamma)\right)
  = \frac{-3g_N^3N\gamma}{1-g_N^2N} \ ,\\
\beta_{g_1} &\eqneq& -\frac{g_1^3}{1-g_1^2C_2(G)}
    \left(3C_2(G)-\sum C(R)(1-\gamma)\right)
  = 3g_1^3N(1-\gamma) \ .\nonumber
\label{eqn:BetaForm}
\end{eqnarray}
The denominator of $\beta_{g_N}$ is zero at $g_N^2N=1$,
but for small couplings it is smooth and positive.

There is a linear relation between the first two equations, so setting
the three $\beta$ functions to zero gives us two conditions on the three
couplings. This yields a fixed line at $\gamma=0$, $g_1=0$.

This fixed line, however,  is not the natural line $g_N=g_1=h$ from which we
want to start the RG flow.
We have already shown that on the natural line $\beta_h$ vanishes up to two
loop order and $\beta_{g_N}$ vanishes up to three loop order.
The form of the $\beta$ functions \eqref{eqn:BetaForm} tells us that
if one $\beta$ function vanishes then so does the other.
It is tempting to speculate that on the natural line both $\beta_h$
and $\beta_{g_N}$ vanish to all orders in perturbation theory.

By an explicit calculation of the $\beta$ functions we can parameterize the
fixed line $\gamma(g_N,g_1,h)|_{g_1=0}=0$ and check whether the natural
line flows to the fixed line in the IR.
The $\beta$ functions of the gauge couplings up to 2-loop order and the
Yukawa coupling up to 1-loop order are
\begin{eqnarray}
\beta_h &=& \frac{6}{N}\frac{h}{(4\pi)^2}\left(N^2h^2-N^2g_N^2-(g_1^2-g_N^2)\right) \ ,\nonumber\\
\beta_{g_N} &=& -12\frac{g_N^3}{(4\pi)^4}\left(N^2h^2-N^2g_N^2-(g_1^2-g_N^2)\right) \ ,
\label{eqn:N1Beta}\\
\beta_{g_1} &=& 3\frac{g_1^3N}{(4\pi)^2}
                -12\frac{g_1^3}{(4\pi)^4}\left(N^2h^2-N^2g_N^2-(g_1^2-g_N^2)\right) \ .\nonumber
\end{eqnarray}
The $\beta$ functions are of the expected form \eqref{eqn:BetaForm}, giving
a consistency check of our calculations.
The first two $\beta$ functions vanish on the two dimensional manifold
defined by the relation
\begin{eqnarray}
h^2=g_N^2+\frac{g_1^2-g_N^2}{N^2} \ .
\end{eqnarray}
The natural line $g_1=g_N=h$ is obviously on this manifold.
The relation between $h, g_N$ on the fixed line $(g_1=0)$ is
\begin{eqnarray}
h^2=(1-\frac{1}{N^2})g_N^2 \ .
\label{eqn:hgRatio}
\end{eqnarray}
In the large $N$ limit the fixed line coincides with the natural line.

\begin{figure}[b]
\begin{center}
\setlength{\unitlength}{0.00016667in}
\begin{picture}(25000,14000)
\put(0,0){\psfig{figure=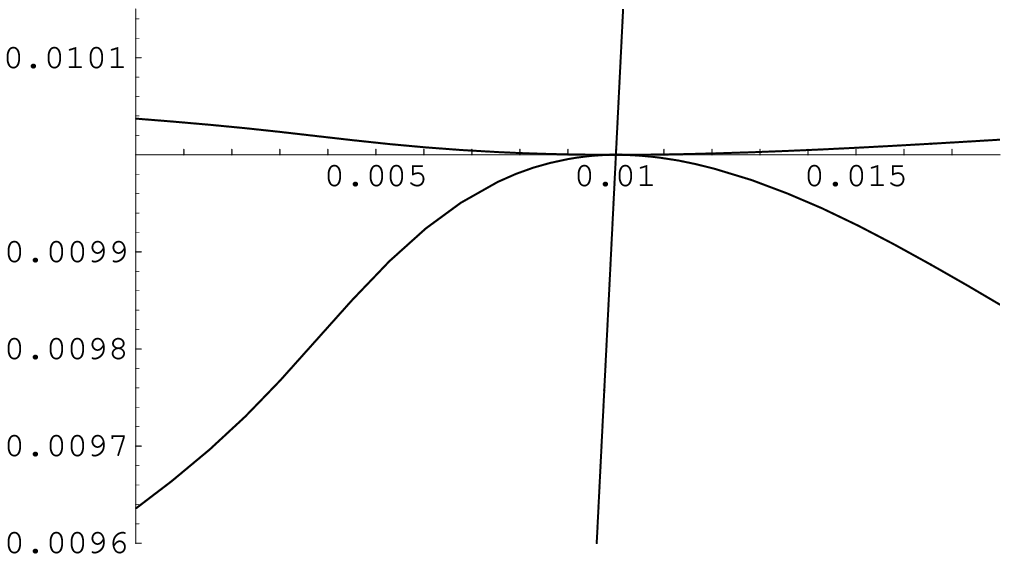}}
\put(14800,13600){$g_1^2$}
\put(23000,10600){$g_N^2$}
\put(24000,9600){$g_1^2$}
\put(24000,6000){$h^2$}
\end{picture}
\end{center}
\isucaption{The RG flow of the $\N=1$ orbifold theory from the natural line
  $g_1^2=g_N^2=h^2=0.01$ to the fixed line in the IR.
  $g_1^2,g_N^2,h^2$ are plotted as a function of $g_1^2$.
  The graph was plotted for $N=5$ and gives the expected
  $h/g_N$ ratio at $g_1=0$.}
\label{fig:N1Flow}
\end{figure}

In figure \ref{fig:N1Flow}
we plot the numerical solution of the $\beta$ functions \eqref{eqn:N1Beta}.
We plot $g_1^2, g_N^2, h^2$ as a function of $g_1^2$
and not as a function of the energy scale $\mu$.
This is permissible because $g_1$ is a monotonic increasing function of
$\mu$ ($\beta_{g_1}$ is positive definite).

The solution demonstrates how the natural line flows to the fixed line
in the IR.
In the IR $(g_1=0)$ we get the expected ratio \eqref{eqn:hgRatio} between
$h$ and $g_N$. This fix line is stable under fluctuations in $g_1,g_N,h$.
This consolidates the assertion that this orbifolded string theory is
described in the IR by an effective field theory on a fixed line.

\subsection{$\N=0$ Orbifolds}
\label{sec:N0}

For the non supersymmetric orbifolds we will focus on the $\Z_2$ orbifold
with weights $(1,1,1,1)$. 
The $\Z_2$ is in the center of $SU(4)_R$, as can be seen from the weights
that do not break the $SU(4)$ symmetry.
We will specify which of the essential results are general and which
are specific to this case.
Applying the $\Z_2$ orbifold to a Type II string theory
reproduces the Type 0 string theory.
The field theory we are describing lives in Type 0B string theory on $N$
dyonic (electric-magnetic) D3 branes \cite{Klebanov:99}.
The theory has the following matter content,
\begin{eqnarray*}
\begin{array}{l|c@{}c@{}c@{}c|c}
 & \multicolumn{1}{c@{\ \otimes}}{SU(N)}
 & \multicolumn{1}{c@{\ \otimes}}{SU(N)}
 & \multicolumn{1}{c@{\ \otimes}}{U(1)}
 & \multicolumn{1}{c|}{U(1)}
 & SU(4) \\
\hline
A^{\mu}_{N_1} & G & 1 & 0 & 0 & 1 \\
A^{\mu}_{N_2} & 1 & G & 0 & 0 & 1 \\
A^{\mu}_{1_1} & 1 & 1 & 0 & 0 & 1 \\
A^{\mu}_{1_2} & 1 & 1 & 0 & 0 & 1 \\
\hline\vphantom{\overline{N}^1}
(\psi^I)^1_2 & N & \overline{N} & \phantom{-}1 & -1 & 4 \\
(\psi^I)^2_1 & \overline{N} & N & -1 & \phantom{-}1 & 4 \\
\hline
\phi^{IJ}_{N_1} & G & 1 & 0 & 0 & 6 \\
\phi^{IJ}_{N_2} & 1 & G & 0 & 0 & 6 \\
\phi^{IJ}_{1_1} & 1 & 1 & 0 & 0 & 6 \\
\phi^{IJ}_{1_2} & 1 & 1 & 0 & 0 & 6
\end{array}
\end{eqnarray*}
where $I,J=1\ldots 4$ are the $SU(4)$ fundamental indices.
The fermions are Weyl spinors and the scalars are real.

Because the scalars are in the adjoint representation there can
only a be Coulomb branch and the gauge group
can maximally break into its Cartan subalgebra $U(1)^{2N}$.
Therefore, there is no spontaneous symmetry breaking of the $U(1)$
gauge fields specified in the table above.
This is not true for all $\N=0$ orbifolds because in general, we can
have scalars in the bifundamental representation,
for example, $\Z_5$ with weights $(1,2,3,4)$ \cite{Kachru:98}.
On the other hand there are $\N=0$ orbifolds with no scalars at all
in the adjoint representation,
for example, $\Z_5$ with weights $(1,1,1,2)$.

The Yukawa terms in the Lagrangian are
\begin{eqnarray}
\frac{Y_N}{2}\sum_{k=1}^2\varepsilon_{IJKL}
\tr\left(\phi^{IJ}_{N_k}(\psi^K)^1_2(\psi^L)^2_1\right) +
\frac{Y_1}{2}\sum_{k=1}^2\varepsilon_{IJKL}
\tr\left(\phi^{IJ}_{1_k}(\psi^K)^1_2(\psi^L)^2_1\right) \ .\ 
\end{eqnarray}
The $U(1)$ scalars couple to the theory only through the $Y_1$
Yukawa coupling.

The quartic terms in the Lagrangian are
\begin{eqnarray}
\frac{\lambda}{4}\tr\left(\
\left[\phi^{IJ}_{N_1},\phi^{KL}_{N_1}\right]^2 +
\left[\phi^{IJ}_{N_2},\phi^{KL}_{N_2}\right]^2
\right) \ .
\label{eqn:quartic}
\end{eqnarray}
This is the classical scalar potential.
The $U(1)$ scalars do not participate in this potential because they are
abelian.

We have a classical moduli space
that can be parametrized by the diagonalized scalar {\it vev} matrices
\begin{eqnarray}
\begin{array}{r@{}c@{}l}
(\phicl[N_k])^{IJ}&=&
  \displaystyle\diag\left((y_k^{IJ})^1,\ldots,(y_k^{IJ})^N\right)
  - \smallfrac{1}{N}\sum_{i=1}^N (y_k^{IJ})^i \cdot\One_{N\times N} \ ,\\
(\phicl[1_k])^{IJ}&=&
  \displaystyle\smallfrac{1}{N}\sum_{i=1}^N (y_k^{IJ})^i\equiv
     \smallfrac{1}{N}\tr(y_k^{IJ}) \ .
\end{array} \s{k=1,2}
\end{eqnarray}

Since there is no supersymmetry, nothing protects the fields from
acquiring a mass by quantum corrections.
There are also new quartic scalar coupling terms that can
appear in the renormalization process \cite{Csaki:99}, for example,
\begin{eqnarray}
\lambda'
\tr\left(\varepsilon_{IJKL}\phi^{IJ}_{N_1}\phi^{KL}_{N_1}\right)
\tr\left(\varepsilon_{IJKL}\phi^{IJ}_{N_2}\phi^{KL}_{N_2}\right) \ .
\label{eqn:newQuartic}
\end{eqnarray}
Those quantum corrections will in general lift the classical moduli.

In order to analyze the behavior of the theory we use the
Coleman-Weinberg effective potential \cite{Coleman:73} as was done in 
\cite{Zarembo:99,Tseytlin:99} for this model without the $U(1)$ factors.
The ``zero loop'' (tree level)  effective potential comes from expanding
the classical potential around the classical {\it vev}s, i.e. setting
$\phi=\phicl+\phi'$.
There are no masses in the original Lagrangian, but in the effective
Lagrangian the fields acquire masses from the scalar {\it vev}s.
The eigenvalues of the mass matrices are
\begin{eqnarray}
\mu_{\tinybox{gauge}}^{ij}(\phicl)^2&=&g_N^2|y^i-y^j|^2 \ ,\nonumber\\
\mu_{\tinybox{scalar}}^{ij}(\phicl)^2&=&\lambda|y^i-y^j|^2 \ ,\\
\mu_{\tinybox{fermion}}^{ij}(\phicl[1],\phicl[2])^2&=&
  \left|Y_N\left(
\left(y_1^i - \smallfrac{1}{N}\tr(y_1)\right) -
\left(y_2^j - \smallfrac{1}{N}\tr(y_2)\right)\right)\right. \nonumber\\
 && \left.{}+Y_1\left(\smallfrac{1}{N}\tr(y_1) -
              \smallfrac{1}{N}\tr(y_2)\right)\right|^2 \ .\nonumber
\end{eqnarray}
The $SU(4)$ indices are implicit,
where $|y|^2=\frac{1}{2}\varepsilon_{IJKL}y^{IJ}y^{KL}$
and $\mu_{\tinybox{fermion}}$ has two $SU(4)$ indices related to the
$SU(4)$ indices of the fermion to which the mass couples.
Therefore, $\mu_{\tinybox{fermion}}$ has also spinor indices.

The one loop effective potential includes the tree level scalar potential
\eqref{eqn:quartic}, renormalization counterterms, and
the one loop correction to the effective potential,
\begin{eqnarray}
V^{\text{1-loop}}_{\text{eff}} &\eqneq& \phantom{{}-{}} 2\sum_{i,j=1}^N\left[
 V\left(\mu_{\tinybox{gauge}}^{ij}(\phicl[1])^2\right) + 
 V\left(\mu_{\tinybox{gauge}}^{ij}(\phicl[2])^2\right)
\right] \nonumber\\
&& {}+ 6\sum_{i,j=1}^N\left[
 V\left(\mu_{\tinybox{scalar}}^{ij}(\phicl[1])^2\right) + 
 V\left(\mu_{\tinybox{scalar}}^{ij}(\phicl[2])^2\right)
\right] \\
&& {}- 8\sum_{i,j=1}^N\left[
 V\left(\mu_{\tinybox{fermion}}^{ij}(\phicl[1],\phicl[2])^2\right) +
 V\left(\mu_{\tinybox{fermion}}^{ij}(\phicl[2],\phicl[1])^2\right)
\right] \ , \nonumber\\
V(\mu^2)&\eqneq&\frac{1}{2}\int \frac{d^4p}{(2\pi)^4}
           \ln\frac{p^2+\smallfrac{1}{2}\mu^2}{p^2} \nonumber\\
 &\eqneq& \frac{2\pi^2}{2(2\pi)^4}\left(
    \frac{1}{4}\Lambda^2\mu^2
    + \frac{1}{16}\mu^4 \left(
      \ln\frac{\smallfrac{1}{2}\mu^2}{\Lambda^2}-\frac{1}{2}\right)
    + O\left(\frac{\mu^6}{\Lambda^2}\right)
  \right) \ .\ 
\end{eqnarray}

The $\Lambda^2$ term provides the one loop correction to the scalar masses.
As can be seen from the effective potential, the scalar mass is corrected 
by a gauge boson loop, a scalar loop and a fermion loop.
The $\Lambda^2$ term is
\begin{eqnarray}
(V^{\text{1-loop}}_{\text{eff}})_{\Lambda^2} = \frac{\Lambda^2}{64\pi^2}
  \sum_{i,j=1}^N
\begin{array}[t]{l}
\displaystyle\vphantom{\sum}
  (2g_N^2+6\lambda)\left(|y_1^i-y_1^j|^2 + |y_2^i-y_2^j|^2\right) \\
\displaystyle\vphantom{\sum}
- 16Y_N^2\left|
  \left(y_1^i - \smallfrac{1}{N}\tr(y_1)\right) -
  \left(y_2^j - \smallfrac{1}{N}\tr(y_2)\right)\right|^2 \\
\displaystyle\vphantom{\sum}
- 16Y_1^2\left|\smallfrac{1}{N}\tr(y_1) -
                  \smallfrac{1}{N}\tr(y_2)\right|^2 \ .
\end{array}
\end{eqnarray}

On the natural line, where all the couplings are equal to $g$ we get
\begin{eqnarray}
(V^{\text{1-loop}}_{\text{eff}})_{\Lambda^2} &=& -\frac{\Lambda^2}{4\pi^2} g^2N^2
 \left|\smallfrac{1}{N}\tr(y_1) - \smallfrac{1}{N}\tr(y_2)\right|^2 \nonumber\\
 &=& \frac{m^2}{2}\left(\phicl[1_1]-\phicl[1_2]\right)^2 \ ,
\label{eqn:U1mass}
\end{eqnarray}
which generates a quadratically divergent tachyonic mass for the diagonal
$U(1)$ scalar.
The mass of the $U(1)$ scalar comes from the fermion loop
and can be set to zero by choosing $Y_1=0$ as was done in \cite{Tseytlin:99}.
Otherwise, we should include a renormalization counterterm for the
scalar mass.
Using this counterterm we can set the renormalized mass to zero.
It would have been much more elegant if there were some underlying mechanism
that naturally sets this scalar mass to zero as discussed in subsection
\ref{sec:Hierarchy}.

The logarithmic term in the effective potential provides the one loop
correction to the quartic coupling $\lambda$.
It is sufficient to compute this term when only one component out of the
${\bf 6}$ of $SU(4)$ is nonzero.
On the natural line the logarithmic term is
\begin{eqnarray}
\left(V^{\text{1-loop}}_{\text{eff}}\right)_{log(\Lambda)} &=& 
-\frac{8g^4}{256\pi^2}\left(
6\left(\tr(\phicl[N_1]^2)-\tr(\phicl[N_2]^2)\right)^2 \right.\nonumber\\
 && \phantom{-\frac{8g^4}{256\pi^2}}
    -8N\left(\tr(\phicl[N_1]^3)-\tr(\phicl[N_2]^3)\right)
      (\phicl[1_1]-\phicl[1_2])
\nonumber\\
 && \phantom{-\frac{8g^4}{256\pi^2}}
    -12N\left(\tr(\phicl[N_1]^2)+\tr(\phicl[N_2]^2)\right)
      (\phicl[1_1]-\phicl[1_2])^2
\nonumber\\
 && \phantom{-\frac{8g^4}{256\pi^2}}\left.{}\vphantom{{1^1}^1}
    -2N^2(\phicl[1_1]-\phicl[1_2])^4
\right) \ .
\label{eqn:Logarithmic}
\end{eqnarray}
There is no $\tr(\phicl[N]^4)$ term. This is a manifestation of the fact
that the one loop $\beta$ function for the original quartic coupling
\eqref{eqn:quartic} vanishes on the natural line.
In models with scalars in the bifundamental representation this vanishing
occurs only if the $U(1)$ factors are taken into account.

The logarithmic term is of the form $(\phicl)^4\ln(\phicl)^2$. The quartic
coupling is the fourth derivative of the effective potential with respect to
the scalar fields, $\lambda'=\frac{d^4 V_{\text{eff}}}{d{\phicl}^4}$.
This derivative diverges at the origin $(\phicl=0)$, requiring a definition of
a renormalized coupling away from the singularity at some arbitrary
renormalization scale $M$,
\begin{eqnarray}
\lambda'=\left.\frac{d^4 V_{\mbox{eff}}}{d{\phicl}^4}\right|_{\phicl=M}
  =C_{\lambda'}+c_{\lambda'}\frac{3g^4}{4\pi^2}\left(
    \ln\frac{\smallfrac{g^2}{2}M^2}{\Lambda} +\frac{11}{3}\right) \ ,
\end{eqnarray}
where $C_{\lambda'}$ is the renormalization counterterm, and
$c_{\lambda'}$ is a constant depending on the $\phicl$ we differentiate
with respect to, calculated in \eqref{eqn:Logarithmic}.
Since $\lambda'$ did not exist in the original Lagrangian we would like
to require $\lambda'=0$.
We are not allowing arbitrary couplings, therefore the renormalization scale
$M$ is the renormalization scale $\mu_N$ defined for the natural line.
Finally, the effective potential for the new quartic couplings is
\begin{eqnarray}
V_{\text{eff}}=c_{\lambda'}\frac{g^4\phicl^4}{32\pi^2}\left(
  \ln\frac{\phicl^2}{\mu_N^2}-\frac{25}{6}\right) \ .
\end{eqnarray}

The first term in \eqref{eqn:Logarithmic} is the term found in
\cite{Tseytlin:99}. This term leads to a repulsive potential between
{\it vev}s of scalars of the same type. For example taking
$y_1^1=-y_1^2=\rho$ leads to a repulsive potential of the form
$\rho^4\ln\rho^2$. Notice also that there are flat directions in
\eqref{eqn:Logarithmic}, for example $y_1^1=y_2^1=-y_1^1=-y_1^2=\rho$.

The other terms in \eqref{eqn:Logarithmic} give a potential to the
diagonal $U(1)$ scalar. Because of the opposite sign this potential is
attractive at short distances. At long distances this potential is repulsive,
but at the scale it becomes so, higher loop terms should be taken
into account.

Assuming that the choice of taking all the masses to zero on the natural
line is consistent, we go ahead and calculate
the explicit $\beta$ functions for the gauge couplings to two loop order
and for the Yukawa couplings to one loop order,
\begin{eqnarray}
\beta_{g_N}&\eqneq&-\frac{g_N^3}{(4\pi)^4}\left(-24N^2g_N^2 - 8(g_1^2-g_N^2)
  + 24N^2Y_N^2 + 24(Y_1^2-Y_N^2)\right) \ ,\nonumber\\
\beta_{g_1}&\eqneq&\frac{11}{3}\frac{g_1^3N}{(4\pi)^2} \nonumber\\
  &&-\frac{g_1^3}{(4\pi)^4}\left(-8N^2g_N^2 - 8(g_1^2-g_N^2)
  + 24N^2Y_N^2 + 24(Y_1^2-Y_N^2)\right) \ ,\\
\beta_{Y_N}&\eqneq&\frac{Y_N}{(4\pi)^2}\left(-6Ng_N^2 - \frac{6}{N}(g_1^2-g_N^2)
  + 6NY_N^2 + \frac{2}{N}(Y_1^2-Y_N^2)\right) \ ,\nonumber\\
\beta_{Y_1}&\eqneq&\frac{Y_1}{(4\pi)^2}\left(-6Ng_N^2 - \frac{6}{N}(g_1^2-g_N^2)
  + 2NY_N^2 + 4NY_1^2 + \frac{2}{N}(Y_1^2-Y_N^2)\right) \ .\nonumber
\label{eqn:N0beta}
\end{eqnarray}
As expected, the $\beta$ functions for $g_N$ and $Y_N$ vanish on the natural
line where all the couplings are the same. Notice that the $\beta$ function
for $Y_1$ also vanishes on the natural line.
Moreover, $Y_1=Y_N\Rightarrow \beta_{Y_1}=\beta_{Y_N}$, meaning that if
the Yukawa couplings start the same at the renormalization scale, they will
stay the same, at least to one loop order.

It would be interesting to look for fixed points where all the $\beta$
functions are zero.
If we take all the $\beta$ functions to the first non vanishing order,
then the only fixed point is the trivial fixed point where all the couplings
are zero. If we use the $\beta_{g_1}$ that we calculated to two loop order
we get several non trivial fixed points. There is one fixed point on
the natural line,
\begin{eqnarray}
\frac{g_N^2}{(4\pi)^2}=
\frac{g_1^2}{(4\pi)^2}=
\frac{Y_N^2}{(4\pi)^2}=
\frac{Y_1^2}{(4\pi)^2}= \frac{11}{48N} \ .
\label{eqn:N0FixPoint}
\end{eqnarray}
This fixed point, however, is inconsistent, because we did not evaluate all
the $\beta$ functions to the same order and ignored the quartic couplings.
The fixed point is also not very interesting because it is a non
stable fixed point.
All the same, the calculations give the order of magnitude of the range of
validity of our calculations. We can trust the lowest order perturbative
expansion as long as $\frac{g^2N}{(4\pi^2)}\ll\frac{11}{48}$.

In figure {\ref{fig:N0Flow}
\begin{figure}[b]
\begin{center}
\setlength{\unitlength}{0.00016667in}
\begin{picture}(32000,13000)
\put(0,0){\psfig{figure=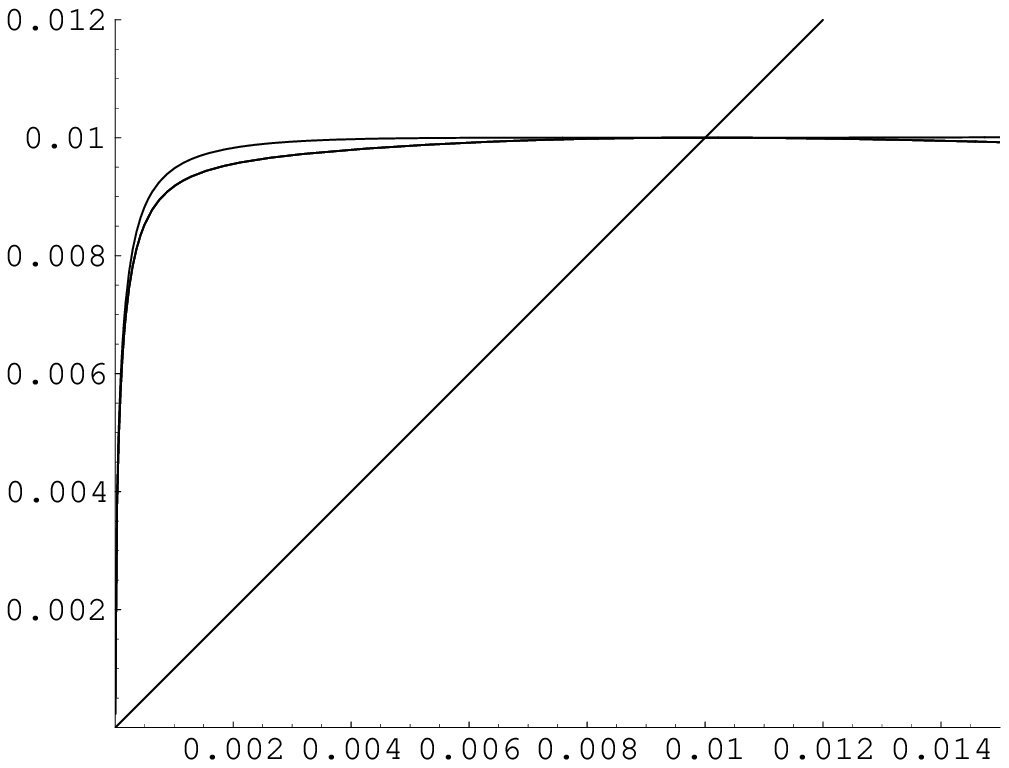,width=2.5in}}
\put(16000,0){\psfig{figure=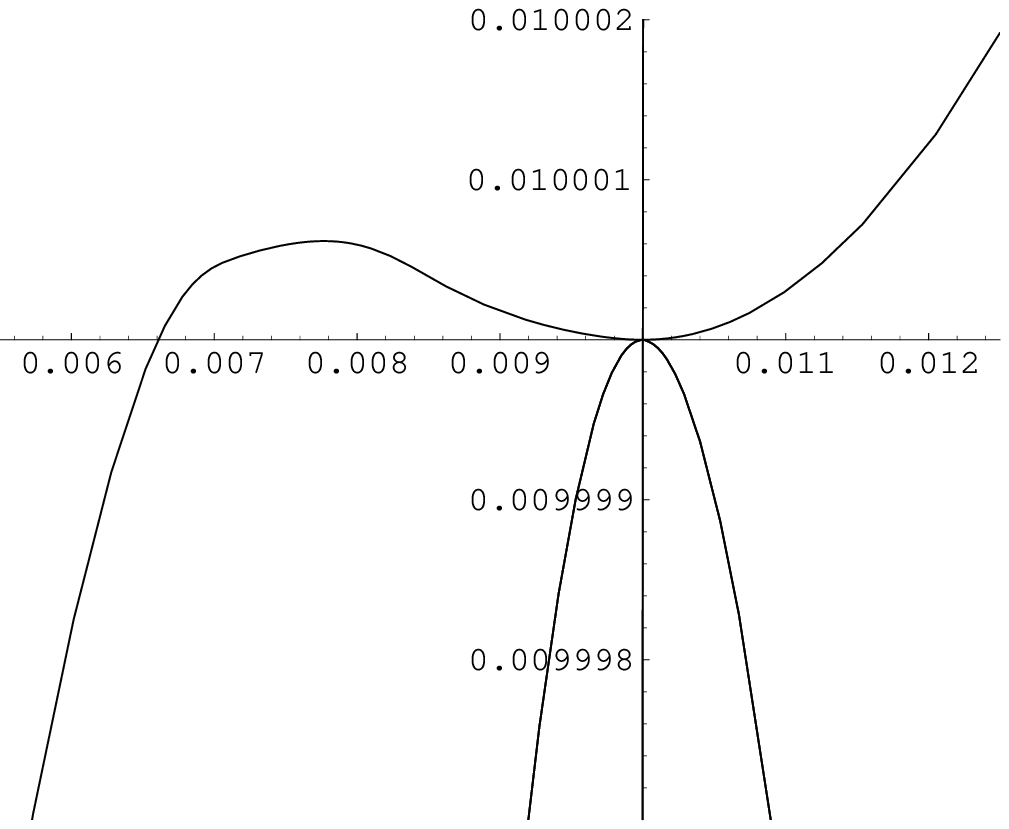,width=2.5in}}
\put(12400,11800){$g_1^2$}
\put(15000,9600){$g_N^2,Y_N^2,Y_1^2$}
\put(15200,800){$g_1^2$}
\put(30800,12200){$g_N^2$}
\put(31200,7200){$g_1^2$}
\put(27800,0){$Y_N^2,Y_1^2$}
\end{picture}
\end{center}
\isucaption{The RG flow of the $\N=0$ theory from the natural
  line $g_1=g_N=Y_1=Y_N=0.01$.
  The graph was plotted for $N=5$.
  In the IR we get a free theory.
  The plot on the right is a zoom on the natural line.
}
\label{fig:N0Flow}
\end{figure}
we plot the numerical solution of the $\beta$ functions.
We start the RG flow from a point on the natural line.
In the IR all the couplings flow to zero. 
The $\beta$ functions \eqref{eqn:N0beta} were calculated assuming that all
the fields are massless, so we can trust them only as long as the
energy scale is larger than the fields masses. Still, if we get an IR free
theory when ignoring the masses, we will get an IR free theory also with
the masses taken into account.

Zooming on the natural point we see that the $SU(N)$ gauge coupling has
a local minimum on the natural line while the Yukawa couplings have
a local maximum. This can be calculated directly from the second derivative
of the couplings (first derivative of the $\beta$ function).

It is possible that higher loop calculations will give more interesting
results than the flow to the trivial fixed point.
This can be seen by analyzing the $\beta$ functions around
the unstable fix point \eqref{eqn:N0FixPoint}.
There seem to be solutions in which all or some of the couplings flow
to infinity in the IR.

\section{Summary and Discussion}
\label{sec:Summary}

We analyzed the finite $N$ limit of orbifold field theories.
For finite $N$ the remnant $U(1)$ factors from the orbifolding procedure
should be considered. 
The $U(1)$ factors seem to cancel the $\frac{1}{N}$ contributions in Feynman
diagrams \eqref{eqn:C2}, at least up to three loops.
This encourages us to study the orbifolded theory on the natural line where
all the couplings are equal.

The simplest case to analyze is of the $\N=2$ orbifolds, but it turns out that
it is too simple. There is no interaction between the different gauge
groups due to the non renormalization theorem.
The $SU(N)$ couplings are finite and the $U(1)$ couplings are IR free.
The theory on the natural line is not finite, but we can easily make it
finite by choosing zero $U(1)$ couplings or by taking the IR limit.

The case of $\N=1$ orbifolds is more interesting because here the $U(1)$
factors do affect the rest of the gauge couplings. Still, the final
conclusions are the same as in the $\N=2$ case. The theory on the natural
line is not conformal, but in the IR limit it flows to a point on the fixed
line.

In the case of $\N=0$ orbifolds all hell breaks loose.
The fact that we get the $\N=0$ theory from the $\N=4$ SCFT leads to
``miraculous'' cancellation of Feynman diagrams up to three loops on
the natural line,
yet this does not seem to be enough. Quantum effects can generate new
terms that did not exist in the original theory, like the mass term
of the $U(1)$ scalar \eqref{eqn:U1mass}.
The theory seems to be inconsistent unless we add a mass term to it,
but adding a mass term is inconsistent with our attempt to look at
the orbifolded theory exactly as it emerges from the orbifolding process.

The scalar mass diverges polynomially, meaning that it is scheme depended.
We hope that there is a regularization scheme in which all the polynomial
divergences cancel out.
The alleged scheme might be defined from the AdS/CFT correspondence
by adding massive fields that correspond to massive open strings
between D3 branes in the orbifolded type IIB string theory.

There is also a problem of renormalization scheme dependence related to our
claims on the cancellation of 3-loop Feynman diagrams.
All the explicit calculations
of the $\beta$ functions we performed where scheme independent, but the 3-loop
Feynman diagrams are scheme dependent. Again, we are not sure in what scheme
our claims are valid, but we assume that such a scheme exists.

Another property of the $\N=0$ theories is that we can no longer assume
that the $U(1)$ factors decouple in the IR, leaving us with an
$SU(N)^{|\Gamma|}$ theory. 
Our analysis shows that, at least for small couplings, the entire
theory is IR free. Moreover, if we start with the same Yukawa couplings
for the $SU(N)$ scalars and the $U(1)$ scalars,
then our calculation shows that at least to one loop order,
they will remain the same.

The five loop bubble diagram \eqref{eqn:5Loops} hindered us from claiming
that the cancellation of the Feynman diagrams continues to all orders.
There are a few hints that this cancellation might survive to all orders,
based on the fact that the $\N=4$ theory is finite to all orders.

The first hint is that the proof that the correlation functions of
orbifold theories coincide
with those of $\N=4$ can be generalized to non planar diagrams.
The proof is valid for abelian orbifolds,
at least as long as all the external legs are on the same face.
In \cite{Bershadsky:98} it was shown, that non planar
diagrams in the orbifold theories are different from the $\N=4$ diagrams,
using as example, the non planar diagram \eqref{eqn:NonPlanar}.
However, in eq. (13) there is a missing $\gamma_3$ factor.
After adding it, there is a
match in the non planar diagram \eqref{eqn:NonPlanar}
between the orbifold and $\N=4$ theories, at least for abelian
orbifolds.

The second hint comes from dividing the Feynman diagrams of each order
into subsets defined by their $N$ dependence.
For example, five loop bubble diagrams can be divided into two subsets,
``calculable'' diagrams with a group factor of $N^4d(G)$ and diagrams with
subleading $N$ terms (like the diagram in \eqref{eqn:5Loops}).

We know that the $\N=4$ theory is finite to all orders independent of
the coupling $g$, meaning that there is a cancellation of the Feynman
diagrams at every order. The finitude of the $\N=4$ theory also does
not depend on $N$ meaning that there is a cancellation of the Feynman 
diagrams in each subgroup defined above.

We can use this cancellation to claim that the five loop diagrams in the
first subset must have zero contribution to orbifold theories.
However, we can not make the same claim for the second subset because
in that subset there are different diagrams with different $N$
dependence.

We have only analyzed orbifolds from the field theory point of view.
It would be interesting to find a manifestation of the $U(1)$ running
coupling constants in string theory orbifolds and in brane configurations. 
The brane configuration for the $\N=2$ theory, for example,
is a set of $|\Gamma|$
NS5 branes on a circle with $N$ D4 branes stretched between them.
The decoupling of the $U(1)$ factors can be seen directly from the brane
configuration \cite{Witten:97}.

In the AdS/CFT correspondence it would be interesting to find string states
corresponding to operators with $U(1)$ factors. The correspondence was
done in \cite{Oz:98} for orbifolds in the large $N$ limit. We do not know
how to generalize it for finite $N$, but it might be possible to use our
knowledge in field theory to gain some insight of the string theory.

In addition to the conformal $SO(4,2)$ symmetry, the original $\N=4$ theory
has an $SL(2,\Z)$ symmetry. It would be interesting to analyze
the effect of the orbifold on this symmetry.
The $SL(2,\Z)$ symmetry acts on the coupling constant and it is not clear
whether it has any meaning in non conformal theories.
In \cite{Bergman:99} it was suggested that $SL(2,\Z)$ is also a symmetry
of the type 0B string. If this is true, then
it would be interesting to investigate the
manifestation of the $SL(2,\Z)$ symmetry in the non supersymmetric
non conformal ``type 0'' field theory.

To summarize, orbifolds of the $\N=4$ field theory give
us non supersymmetric non conformal theories with very interesting features.
The exquisite cancellation of the vacuum bubble diagrams in those theories
up to at least four loop order suggests that
the cosmological constant is very small in those theories.
Consequently, the cosmological constant problem could be solved in
the orbifolded theories.
We do not claim that the cosmological constant vanishes completely in
those theories because the running of the $U(1)$ factors stimulate the
running of the bubble diagrams.

For finite $N$ the $U(1)$ factors also behave as soft symmetry breaking terms
of the conformal symmetry in the sense that the flow of the $U(1)$
couplings induces the flow of the other couplings.
The ``soft breaking'' parameter is $\frac{1}{N}$.
The conformal symmetry is broken explicitly,
but it is broken by terms that occur naturally in the orbifold process.
We propose this soft symmetry breaking term for solving the hierarchy problem
as suggested in \cite{Frampton:99}.

\subsection*{Acknowledgments}

I would like to thank Amihay Hanany, Vadim Kaplunovsky, Shimon Yankielowicz
and especially Jacob Sonnenschein for many useful discussions.
I would also like to thank Angel Uranga for pointing out to me the
anomalies in the chiral orbifolds.

\bibliography{hep}
\bibliographystyle{utcaps}

\end{document}